# Battery-like Supercapacitors from Vertically Aligned Carbon Nanofibers Coated Diamond: Design and Demonstrator

Siyu Yu,[a] Nianjun Yang,[a*] Michael Vogel,[a] Soumen Mandal,[b] Oliver A. Williams,[b] Siyu Jiang,[c] Holger Schönherr,[c] Bing Yang,[d] and Xin Jiang[a,d*]

[a] Institute of Materials Engineering, University of Siegen, 57076 Siegen, Germany

[b] School of Physics and Astronomy, Cardiff University, Cardiff CF24 3AA, UK

[c] Physical Chemistry I, Department of Chemistry and Biology, University of Siegen, 57076 Siegen, Germany

[d] Shenyang National Laboratory for Materials Science, Institute of Metal Research (IMR), Chinese Academy of Sciences (CAS), No.72 Wenhua Road, Shenyang 110016 China

E-mail: nianjun.yang@uni-siegen.de, xin.jiang@uni-siegen.de

## Abstract

Battery-like supercapacitors feature high power and energy densities as well as long-term capacitance retention. The utilized capacitor electrodes are thus better to have large surface areas, high conductivity, high stability, and importantly be of binder free. Herein, vertically aligned carbon nanofibers (CNFs) coated boron-doped diamonds (BDD) are employed as the capacitor electrodes to construct battery-like supercapacitors. Grown via a thermal chemical vapor deposition technique, these CNFs/BDD hybrid films are binder free and own porous structures, resulting in large surface areas. Meanwhile, the containment of graphene layers and copper metal catalysts inside CNFs/BDD leads to their high conductivity. Electric double layer capacitors (EDLCs) and pseudocapacitors (PCs) are then constructed in the inert electrolyte (1.0 M $H_2SO_4$ solution) and in the redox-active electrolyte (1.0 M $Na_2SO_4$ + 0.05 M $Fe(CN)_6^{3-/4-}$), respectively. For assembled two-electrode symmetrical supercapacitor devices, the capacitances of EDLC and PC devices reach 30 and 48 mF cm$^{-2}$ at 10 mV s$^{-1}$, respectively. They remain constant even after 10 000 cycles. The power densities are 27.3 kW kg$^{-1}$ and 25.3 kW kg$^{-1}$ for EDLC and PC devices, together with their energy densities of 22.9 Wh kg$^{-1}$ and 44.1 Wh $kg^{-1}$, respectively. The performance of formed EDLC and PC devices is comparable to market-available batteries. Therefore, the vertically aligned CNFs/BDD hybrid film is a suitable capacitor electrode material to construct high-performance battery-like and industry-orientated supercapacitors for flexible power devices.

## Introduction

Battery-like supercapacitors refer to those electrochemical capacitors (ECs) which own the features of both ECs (e.g., high power density, *P*) and batteries (e.g., high energy density, *E*) as well as high capacitance (*C*) retention. These ECs then meet with the demands for powering future multifunctional electronics, hybrid electric vehicles, and industrial equipment.[1] Nowadays, construction of such battery-like supercapacitors is the core of EC researches for both academic scientists and industrial engineers.

To fabricate battery-like supercapacitors, the first issue to be considered is the choice of suitable capacitor electrodes. Besides large surface areas, it is more essential for them to facilitate high electron and ion mobility. To grow capacitor electrodes owning such properties, several strategies have been proposed through[2]: 1) improving the conductivity of the electrodes; 2) employing nano-sized electrode materials to reduce the diffusion length and meanwhile to enhance the surface areas; 3) utilizing three-dimensional (3D) materials to realize ion diffusion in multiple directions (e.g., porous materials); 4) reducing diffusion activation energy of the ions. In these studies, various carbon materials with different hybridization of atomic orbitals ($sp^2$ and $sp^3$), altered allotrope forms of carbon (e.g. fullerenes, nanotubes, graphene and diamond etc.), and various dimensionalities (e.g., 0 to 3D) have been intensively investigated.[3] For most ECs in these approaches, these carbon materials are required to be pressed or coated on an electrode supporter (a current collector) with organic binders (e.g., polytetrafluoroethylene[4]). Due to the low conductivity and relatively poor stability of these organic binders, both electron mobility inside the formed capacitor electrodes and ion diffusion on the surface of these capacitor electrodes are partially hindered. In most cases, the expected performance of fabricated ECs has not been delivered. Therefore, novel capacitor electrodes, namely those bind-free carbon capacitor electrodes, are highly needed for the construction of battery-like supercapacitors.

The second crucial issue for the construction of battery-like supercapacitors is the selection of the electrolytes. For example, inert electrolytes are widely employed for the formation of electrical double layer capacitors (EDLCs). In these solutions, redox species coated conduc-tive substrates with polymer and metal oxides are also employed as the capacitor electrodes to produce pseudocapacitors (PCs).[5]. An alternative but more efficient approach to construct PCs is to introduce soluble redox species into the electrolyte.[6] One of the main advantages of this novel approach is that the amount of redox species, which determines the charge storage capacity, is easy to be controlled. However, the contribution of soluble redox electrolytes into the performance of ECs has not been clearly clarified in the literature.[7] Therefore, the development of battery-like supercapacitors is still highly required in both inert and redox electrolytes from the viewpoints of fundamental and practical aspects.

Herein, we propose the growth of a novel binder-free carbon capacitor electrode as well as its employment for the construction of battery-like supercapacitors, including EDLCs in inert electrolytes and PCs in redox electrolytes. The binder-free carbon capacitor electrode is constructed using boron doped diamond (BDD) as the electrode supporter and vertically aligned carbon nanofibers (CNFs) as the active electrode material. The CNFs are directly grown on BDD with a thermal chemical vapor deposition process (TCVD). Namely, such a TCVD process eliminates the need to use low conductive organic binders. Moreover, in this way the CNFs are covalently bonded to BDD via a stable C-C chemistry. In other words, CNFs on diamond have good adhesion of the films, increased conductivity, and long-termed stability. Furthermore, the CNFs feature high electrical conductivity, large specific surface area, good chemical stability, and 3D porous structures.[8] Abundant diffusion channels and plenty of active sites are thus expected to be available for our capacitor electrode. Therefore, the hybrid structure of CNFs/BDD hybrid films will exhibit integrated properties of both carbon materials. The issues mentioned above for the construction of battery-like supercapacitors are promising to be solved.

In this contribution, we first present the TCVD growth of vertically aligned CNFs on BDD. For the growth of CNFs on BDD, $C_2H_2$ was employed as the reaction gas and the copper (Cu) thin film sputtered with a physical vapor deposition (PVD) device acted as the catalyst. The morphology, chemical structure, wettability of these CNFs films are then characterized using SEM, TEM, XPS, Raman, and Contact angel measurements. The construction of battery-like

EDLC and PC supercapacitor devices using CNFs/BDD hybrid film capacitor electrodes are then shown. The performance of developed EDLCs in the aqueous electrolyte (1.0 M $H_2SO_4$ solution) as well as PCs in the redox electrolyte (1.0 M $Na_2SO_4$ + 0.05 M $Fe(CN)_6^{3-/4-}$) are evaluated and compared using both three- and two-electrode systems. A stand-alone supercapacitor device is demonstrated.

## Experimental Section

### Electrode materials

Figure S1 illustrates schematically the steps for the growth of CNFs/BDD hybrid films using a TCVD technique. First, BDD films were grown on silicon wafers using microwave plasma assisted chemical vapor deposition (MWCVD) technique.[9] Then, the coating of BDD with copper films was carried out on a physical vapor deposition (PVD) device. The high purity (99.999%) copper (4 inch) was used as the target. Such a PVD was equipped with a turbo molecular pump and its base pressure was lower than $5\times10^{-6}$ mbar. Prior to coating, a pre-sputtering of the target for 10 min with a closed shutter was applied to clean the target. Other RF magnetron sputtering conditions were: room temperature, argon atmosphere, argon gas flow of 50 sccm, a pressure of 3.5 – $4.5\times10^{-3}$ mbar. The thickness of copper films on BDD was varied through altering the sputtering times ($t_{Cu,s}$). In this study, $t_{Cu,s}$ was varied from 15, 30, 60, 90, to 120 s. After that, copper coated BDD films were introduced in the center of a quartz tube in a TCVD device. To grow CNFs, these sputtered copper nano-films were utilized as the catalyst. At a pressure of about $5\times10^{-2}$ mbar, the tube was heated to 250 °C with a heating rate of 5 °C $min^{-1}$. Subsequently, the reaction gas of $C_2H_2$ was filled into the tube with a pressure of 500 mbar. The growth times were varied from few minutes to few hours. After the growth, the reactor was rapidly evacuated. Once the pressure in the tube was lower than $5\times10^{-2}$ mbar, the carbonization of CNFs was carried out. The carbonization temperature was 800 °C and the time applied was 60 min. As control experiments, some copper nano-films were annealed before the growth of CNFs in the TCVD device under the conditions of an annealing temperature of 500 °C, a pressure of about $5\times10^{-2}$ mbar, and an anneal-ing time for 60 min. To change the wettability of CNFs/BDD hybrid films, they were immersed in a mixture of $H_2SO_4$ and $HNO_3$ (v/v = 3:1) for 30 min, cleaned with deionized water, and finally dried in a $N_2$ atmosphere before electrochemical experiments.

**Characterization**

The surface and cross section morphologies of CNFs/BDD hybrid films were investigated with a field emission scanning electron microscopy (FESEM, Zeiss ultra55, Germany). The Raman spectra of the films were recorded on a homemade micro Raman configuration with a 532 nm laser. An X-ray photoelectron spectroscopy (XPS, Surface Science Instruments, SSX-100 S-probe photoelectron spectrometer, USA) with an Al Kα radiation of 200 W was used to characterize the elemental composition of the films. Static contact angle measurements were carried out on an OCA 15plus instrument (Data Physics Instruments GmbH, Filderstadt, Germany) with Milli-Q water. Transmission electron microscopy (TEM) images of carbon nanofibers were acquired using a Tecnai G2 F30.

**Electrochemical measurements**

Electrochemical measurements were conducted on a CHI660E Potentiostat / Galvanostat (Shanghai Chenhua Inc., China). A standard three-electrode cell was applied where a CNFs/BDD hybrid film acted as the working electrode, an Ag/AgCl (3 M KCl) electrode as the reference electrode, and a coiled Pt wire as the counter electrode. For a two-electrode symmetrical supercapacitor device, two CNFs/BDD hybrid films behaved as the capacitor electrodes and a 50 μm thick Nafion film as the separator. The geometric area of a CNFs/BDD hybrid capacitor electrode was 0.05 $cm^2$. For EDLCs, the electrolyte was 1.0 M $H_2SO_4$. For PCs, the electrolyte was 1.0 M $Na_2SO_4$ containing 0.05 M $K_3Fe(CN)_6/K_4Fe(CN)_6$. The cyclic voltammograms (CVs) were recorded at scan rates ranging from 10 to 100 mV $s^{-1}$. The galvanostatic charge/discharge (GCD) curves for EDLCs and PCs were obtained at different current densities. The specific capacitances ($C$, F $cm^{-2}$), energy density ($E$, W h $kg^{-1}$), and power density ($P$, W $kg^{-1}$) of ECs were calculated according to the reported methods.[10]

## Results and Discussion

### Design of CNFs/BDD Capacitor Electrodes

A best capacitor electrode features high electrical conductivity, large specific surface area, good chemical stability, and 3D porous structure. To fabricate such a capacitor electrode, the growth of CNFs on BDD was first optimized. It has been shown that in a TCVD technique the growth (e.g., size, length, density) of CNFs is actually determined by the thickness of a catalyst (in our case, the Cu films) .[11] The SEM images of copper films sputtered at BDD films with $t_{Cu,s}$ of 15, 30, 60, 90, and 120s were thus recorded. Figure 1a shows one typical SEM image of a Cu film with a $t_{Cu,s}$ of 60 s. The SEM images with $t_{Cu,s}$ of 15, 30, 90, and 120 s are shown in Figure S2. In all these images, BDD films are fully covered with Cu. The thicknesses of Cu films vary from a dozen to tens of nanometers, increasing as a function of $t_{Cu,s}$. The crystal boundaries of BDD can be clearly observed with $t_{Cu,s}$ from 15 to 90 s, but become indistinct once a $t_{Cu,s}$ is up to 120 s. These Cu films were then applied as the catalyst for the growth of CNFs. The surface morphologies of these CNFs were checked using SEM. The top and side views of as-grown CNFs/BDD hybrid films for a $t_{Cu,s}$ = 60 s are shown in Figure 1b. Those for $t_{Cu,s}$ = 15, 30, 90, 120 s are shown in Figure S3. For a $t_{Cu,s}$ of 15 s, the growth of CNFs is random, while CNFs are quasi vertically aligned when a $t_{Cu,s}$ is up to 30 s. When $t_{Cu,s}$ is higher than 60s, vertically aligned CNFs with much denser arrangement are acquired. The thicknesses of CNFs (insets in SEM images) are measured to be about 2.5, 2.8, 3.6, 4.3, and 5.0 µm for $t_{Cu,s}$ of 15, 30, 60, 90, and 120 s, respectively. Interestingly, all CNFs films exhibit 3D porous properties. Lots of channels or pores exist between the CNFs, even for the CNFs with $t_{Cu,s}$ longer than 60 s. This is partially because the growth direction of CNFs is perpendicular to the substrate. The surface morphology of CNFs films is closely dependent on that of the substrate. For instance, the surface characteristic of the CNFs film for a $t_{Cu,s}$ of 60 s (Figure 1b) shows actually the morphology of BDD surface. At some irregular crystal boundaries of BDD, tilted CNFs are then obtained and pores are generated. Such a statement is further supported by the control experiments using a smooth Si substrate as the supporter for the CNFs growth. In this case, only smooth surface of the CNFs films was obtained under identified conditions (Figure S4).

To figure out the effect of the morphology of Cu catalysts on the growth of CNFs on BDD, the growth of CNFs using Cu particles as catalyst was further tested (Supporting Information). These Cu particles were generated via annealing the Cu nano-films at 500 °C for 60 min. Vertically aligned CNFs are gained with a $t_{Cu,s}$ of 90 and 120 s. Interestingly, the thickness of CNFs films are slightly smaller than those CNFs films obtained without annealing of the Cu films (WOA). Therefore, the copper films WOA were used throughout our studies.

Obviously, altering the growing time leads to the growth of CNFs with various lengths. In a case study, Cu films sputtered with $t_{Cu,s}$ of 60 s WOA were applied for the growth of CNFs using different growth times. The SEM images of CNFs grown using a reaction time of 30, 60, 90, and 120 min are shown in Figure 1b, Figure 1c, Figure S7a, and Figure S7b, respectively. As expected, the surface morphologies of these CNFs/BDD hybrid films are almost identical. The lengths of CNFs are about 1.8, 3.6, 5.5, 7.2 - 8.0 μm for a growth time of 30, 60, 90, and 120 min, respectively. Consequently, longer CNFs are possible to be attained once a longer growth time is applied or a high $C_2H_2$ concentration is used.

Prior to employment of CNFs/BDD hybrid films as the capacitor electrodes, their wettability was checked. The as-grown CNFs/BDD hybrid films were not water-wettable. This is characteristics for the CNFs. Wet-chemical treatment was then applied to these CNFs via immersing them in a mixture of $H_2SO_4$ and $HNO_3$ (v/v=3:1) for 30 minutes. As a case study, the wettability of CNFs with $t_{Cu,s}$ of 60 s WOA (Figure 1b) was examined. The contact angel of water on the as-grown CNFs was detected to be 110.4° ± 1° (Figure S8a), revealing its hydrophobic nature. Its XPS survey spectrum (Figure S8c) exhibits the C 1s (98.08 %) and O 1s (1.92 %) signals at 284 and 532 eV, respectively. After such a wet-chemical treatment, the contact angel of water on the treated CNFs surface was changed to be 24° ± 0.6° (Figure S8b). Accordingly, the content of O 1s estimated from its XPS survey spectrum, rises to 11.61 % (Figure S8d). Small amount of S 1s (0.35 %) is observed at 168 eV, probably due to the residual acid on the surface. Therefore, the wet-chemical treatment of CNFs enhances the degree of hydrophilic terminations, leading to significantly improved wettability of CNFs/BDD hybrid films in aqueous solution. Such a wet-chemical treatment was then always applied for the CNFs used for electrochemical experiments.

To optimize the CNFs/BDD hybrid films for the later construction of both EDLCs and PCs, cyclic voltammograms (CVs) of CNFs/BDD hybrid films with $t_{Cu,s}$ ranging from 15 to 120 s with (WA) and without (WOA) Cu annealing were recorded in 1.0 M $H_2SO_4$ at a scan rate of 100 mV $s^{-1}$. The growth time of CNFs was 60 min. For these tests, a three-electrode system was used. Figure S9a shows two typical CVs for CNFs/BDD WA and WOA films with $t_{Cu,s}$ of 60 s. The capacitive current of the CNFs/BDD hybrid film WOA is larger, an indication of a higher capacitance. Figure S9b summarizes the calculated capacitances of all the CNFs/BDD hybrid films estimated from their corresponded CVs as a function of $t_{Cu,s}$. In the case of CNFs/BDD hybrid films WA, almost no variation of capacitances is observed when different $t_{Cu,s}$ is applied. These nearly unchanged capacitances suggest the constant electrode areas of these CNFs. By using CNFs/BDD hybrid films WOA, the magnitude of the capacitance improves with an increase of $t_{Cu,s}$ up to 60 s and then remains almost constant until 90 s. Further increase of $t_{Cu,s}$ longer than 90 s leads on the contrary to a decrease of the magnitude of the capacitance. Again, the change of the capacitance reflects directly the variation of surface areas of the formed CNFs/BDD hybrid films WOA. For $t_{Cu,s}$ in the range from 60 to 90 s, the highest capacitance (about 36 mF $cm^{-2}$) is achieved. For further capacitance investigation, CNFs/BDD hybrid films WOA grown with a relatively shorter copper sputter time (e.g., $t_{Cu,s}$ = 60 s) were chosen as EC capacitor electrodes.

To reveal the structures of these CNFs, they were further examined with TEM and Raman. Figure 2a shows one representative TEM image of a CNF, of which size is about 200 nm. The black triangle inside the CNF is a Cu catalyst. Its selected area electron diffraction (SAED) pattern of the [110] zone axis is shown in the inset of Figure 2a, proving good crystallinity of a Cu catalyst. Figure 2b shows the high magnification of the interface between Cu catalyst and a CNF. One can see that graphite is formed around the Cu catalyst. Lattice fringes with a distance of about 0.34 nm of the CNF is seen from the high resolution TEM (HRTEM) image of Figure 2c, confirming the formation of graphene-like layers around the Cu catalyst[12]. In the region far from Cu catalyst characteristic of amorphous carbon is detected, based on the HRTEM image and the fast Fourier transformation (FFT) image showed in Figure 2d. In addition, a large number of the CNFs are noticed outside this area, indicating that graphitization only occurs in the region around Cu catalyst.

Further recorded Raman spectrum of the CNFs/BDD hybrid film (Figure S10) reveals two major Raman bands: D band at ~1370 $cm^{-1}$ and G band at ~1650 $cm^{-1}$. The D band is attributed to the disordered structure or defects of graphitic sheets. The G band indicates the crystalline graphitic structure.[13] The ratio of the intensity of the G band ($I_G$) to that of D band ($I_D$), calculated using Gaussian fitting of the two peaks, is known to be proportional to the graphitization degree of carbon materials. In our case, $I_G/I_D$ was estimated to be about 0.42, confirming the large amount of amorphous phase and relatively few ordered graphite crystallites in the CNFs.[12] The broad peak observed at around 2800 $cm^{-1}$ is probably related to the 2D band of graphite layers.

The existence of graphite and Cu inside the CNFs improves the electrical conductivity of the formed material. A more detailed study of the conductivity using scanning tunneling microscope (STM) is still under investigation. Together with its large surface area and a porous structure, CNFs/BDD hybrid films with $t_{Cu,s}$ of 60 s WOA are expected to facilitate ion transfer in the solution and electron mobility in the interface of a CNFs/BDD electrode.

**Performance of CNFs/BDD ECs**

**Capacitance**

The employment of CNFs/BDD hybrid films with $t_{Cu,s}$ of 60 s WOA to fabricate ECs and their capacitances were first studied using a three-electrode system. For the construction of EDLCs, its CVs in 1.0 M $H_2SO_4$ were recorded within the potential range of 0 - 1.0 V at different scan rates (Figure S11a). The CVs are nearly rectangular, indicating ideal EDLC behavior. The slight deviation of the CVs from rectangular shape is due to the altered charge transfer resistance ($R_{ct}$) between CNFs pores/electrolyte interfaces and over-potential.[14] The estimated capacitance is 36.4, 48.2, 80.1, and 116.3 mF $cm^{-2}$ at the scan rate of 100, 50, 20, and 10 mV $s^{-1}$, respectively. Figure S11b presents the galvanostatic charge/discharge (GCD) curves at current densities ranging from 2 to 20 mA cm$^{-2}$. At high current densities, the curves are almost symmetrical, demonstrating the high reversibility of this EDLC. The IR drop observed at the scan rate of 2 mA cm$^{-2}$ is probably due to the internal resistance of CNFs films and $R_{ct}$ caused mainly by diffusion kinetics of the ions. The calculated capacitance is 17.6, 27.7, 56.6, and 137.9 mF cm$^{-2}$ at the current density of 20, 10, 5, and 2 mA cm$^{-2}$, respectively.

The CNFs/BDD PCs were then fabricated by introducing redox species (here 0.05 M $Fe(CN)_6^{3-/4-}$) into 1.0 M $Na_2SO_4$ aqueous solution. Its CVs were recorded in a potential window of -0.2 – 0.8 V at different scan rates (Figure S11c). At all scan rates, the CV curves show a pair of redox waves corresponding to redox reactions of $Fe(CN)_6^{3-/4-}$. The peak potential separation ($\Delta E_p$) is relatively small (e.g., $\Delta E_p$ = 96 mV at a scan rate of 10 mV s$^{-1}$). Moreover, the anodic peak currents are identical to the cathodic ones, indicating the excellent reversibility of these PCs. The evaluated capacitance is 35.0, 53. 5, 94.8, and 136.8 mF cm$^{-2}$ at the scan rate of 100, 50, 20, and 10 mV s$^{-1}$, respectively. The related GCD curves are shown in Figure S11d. All these recorded curves show nonlinear behavior with plateaus, relating to redox reactions of $Fe(CN)_6^{3-/4-}$. The times for the charge and discharge processes are almost identical at all current densities, demonstrating the perfect reversibility of this PC. The estimated capacitance is 14.5, 34.7, 84.6, and 232.0 mF cm$^{-2}$ at the current density of 20, 10, 5, and 2 mA cm$^{-2}$, respectively.

At high scan rates (e.g. 100 mV s$^{-1}$) and high current densities (e.g. 20 mA cm$^{-2}$), the capacitances of a PC are slightly smaller than those of an EDLC. Besides two different charge storage mechanisms (namely charge accumulation for EDLCs, charge transfer and accumulation for PCs,) there exist additional aspects. First, different supporting electrolytes are employed for these ECs. Namely, the supporting electrolyte used for PCs is $Na_2SO_4$, while for EDLCs $H_2SO_4$ is applied as the electrolyte. To clarify the effect of these electrolytes, the CVs of an EDLC in 1.0 M $H_2SO_4$ and in 1.0 M $Na_2SO_4$ were recorded at the scan rate of 100 mV s$^{-1}$ (Figure S12). The capacitive current obtained in $Na_2SO_4$ is much smaller, resulting in a low capacitance of 8.3 mF cm$^{-2}$. The difference can be intercepted with altered conductivity of the electrolytes. For the electrolyte of 1.0 M $H_2SO_4$, the conductivity is 1000 mS cm$^{-1}$. For the electrolyte 1.0 M $Na_2SO_4$, it is 80 mS $cm^{-1}$.[15] Moreover, the size of $Na^+$ ions in hydrated state is larger than that of hydrated protons. The mobility of $Na^+$ ions in the solution is slower. Their accessibility and accumulation to the pores of the densely packed CNFs are harder. If one compares the capacitance of the EDLC using $Na_2SO_4$ to that of a PC, the capacitance of a PC is enlarged for more than 4 times even at high scan rates (100 mV $s^{-1}$). Such enhancement of the capacitance is similar or even better than those reported by using other porous carbon materials.[16] Second, the inferior behavior of the PC at high scan rates or current densities can be attributed to kinetically unfavorable diffusion of ions inside the narrow pores because of slow $Na^+$ ionic motion and low conductivity of the electrolyte, leading to loss of the full contribution of active surface area of a CNFs film.

To figure out clearly the contribution of CNFs into the construction of these ECs, the capacitance of the fabricated EDLC is compared with that of diamond nanostructures related EDLCs as well as with that of hydride films of BDD with other carbon materials (Figure S13). Clearly, BDD nanostructures (e.g., honeycomb diamond[17], porous diamond[18], and diamond networks[10a, 19], etc.) own enhanced surface areas and thus exhibit significantly improved capacitance in comparison to pure BDD. The hydride films of BDD with other carbon materials (e.g., carbon nanotube[20], carbon fiber[21], etc.) also deliver larger capacitance than BDD EDLCs. This is intercepted partially due to further increased surface areas, partially due to the capacitance addition from other carbon materials. Compared with these values, the capacitance of our CNFs/BDD EDLC is the highest. Moreover, the comparison of the capacitances of different diamond PCs is shown in Figure S13. The capacitance of the CNFs/BDD PC reaches the highest value of 232.0 mF $cm^{-2}$. It is much larger than that of metal oxide (e.g. $Ni(OH)_2$[22]) or conducting polymer (e.g. PEDOT[23]) and diamond nanowires based PCs, as well as that obtained when diamond network was utilized as the capacitor electrode and redox solution as the electrolyte[10a]. In summary, the large surface area, the improved electrical conductivity, and the unique porous structure of these vertically aligned CNFs on BDD lead to the high-performance of these EDLCs and PCs.

The effect of the growth time of CNFs on the capacitances of CNFs/BDD hybrid films were also examined in 1.0 M $H_2SO_4$ using a three-electrode system. The CVs of CNFs with a growth time of 30, 90, and 120 min were recorded at a scan rate of 100 mV $s^{-1}$ (Figure S14a). The capacitances calculated from the related CVs are listed in Figure S14b. The magnitude of the capacitance improves almost linearly with an increase of growing time. This is because of the nearly linear enhancement of CNFs length, namely its surface area. In other words, the capacitance of these CNFs/BDD ECs is possible to be further improved as required only through applying longer TCVD growth times.

For practical applications of these ECs, a two-electrode symmetrical supercapacitor device was assembled using two CNFs/BDD hybrid films as the capacitor electrodes. These films were grown with a $t_{Cu,s}$ of 60 s WOA. The performance of as-fabricated ECs was first investigated in 1.0 M $H_2SO_4$ aqueous solution. Figure 3a shows the CV curves recorded at the scan rates of 100, 50, 20, and 10 mV $s^{-1}$ and with a cell voltage of 1.0 V. Similar as the results obtained with a three-electrode system, CVs show nearly rectangular shape at low scan rates, revealing a good double layer capacitive behavior. The capacitance was calculated to be 7.7, 11.7, 21.6, and 30.4 mF $cm^{-2}$ at the scan rate of 100, 50, 20, and 10 mV $s^{-1}$, respectively.

The related GCD curves of the EDLC device at the current densities from 1 to 20 mA $cm^{-2}$ are presented in Figure 3b. The curves are almost symmetric at high current densities, demonstrating the high reversibility of the EC. When lower current densities (here smaller than 2 mA $cm^{-2}$) are applied, asymmetrical curves are acquired, due to the charge transfer resistance at the electrode and electrolyte interfaces. The times required for charge and discharge process are almost equivalent, indicating a high columbic efficiency of the EDLC. The capacitance was calculated to be 2.1, 4.0, 6.8, 19.6, and 34.0 mF $cm^{-2}$ at the current density of 20, 10, 5, 2, and 1 mA $cm^{-2}$, respectively.

Considering that fast charge and discharge process is required for ECs, CV measurements at higher scan rates up to 3 V $s^{-1}$ were carried out. The related CVs are displayed in Figure S15a. Deformation of the CV curves is noticed at these high scan rates. This is quite normal due to insufficient time for ion adsorption/desorption and diffusion into the inner pores. In this context, the capacitance decreases with an increase of the scan rate, as shown in Figure S15b.

Consequently, these CNFs/BDD EDLCs are possible to be applied for fast charge-discharging processes.

On the other hand, the capacitance of a PC device was then evaluated in a cell voltage of -0.5 – 0.5 V in the redox electrolyte of 0.05 M $Fe(CN)_6^{3-/4-}$ + 1.0 M $Na_2SO_4$. At all scan rate, the CV curves (Figure 4a) show a pair of well-defined peaks, relating to the redox reaction of $Fe(CN)_6^{3-/4-}$. In addition, the charges stored on the electrodes during charge and removed during discharge are nearly identical, integrated from the anodic and cathodic cycles, respectively. The calculated capacitance is to be 15.7, 23.3, 36.1, and 48.1 mF $cm^{-2}$ at the scan rate of 100, 50, 20, and 10 mV $s^{-1}$, respectively.

Figure 4b presents the GCD curves of this PC device at different charge/discharge current densities. The plateaus in the curves are relative to the redox reactions of $Fe(CN)_6^{3-/4-}$. Almost same times are required for charge and discharge process even at a low current density of 2 mA $cm^{-2}$. The estimated columbic efficiency is about 100%, confirming the excellent reversibility of such a PC device. The calculated capacitance is 6.4, 12.8, 25.6, and 65.4 mF $cm^{-2}$ at the current density of 20, 10, 5, and 2 mA $cm^{-2}$, respectively.

Interestingly, the capacitance of a PC device is several times larger in comparison to that of an EDLC device. The reason can be ascribed to the fast and more efficient process of charge accumulation on the electrode interface, namely a rapid faradaic reactions of $Fe(CN)_6^{3-/4-}$ redox couples ($[Fe(CN)_6]^{3-} + e^- \leftrightarrow [Fe(CN)_6]^{4-}$) on the CNFs films. This reaction contributes additional or extra pseudocapacitance to the ECs.[10a] In short, both fabricated CNFs/BDD EDLC and PC devices feature the characteristics of battery-like supercapacitors, here a large capacitance.

**Capacitance retention**

The cycling stability of a CNFs/BDD EDLC device was further examined using GCD technique at the current density of 5 mA $cm^{-2}$. After 10 000 cycles, the capacitance remains unchanged (Figure 3c). The morphologies of both electrodes after the lifetime test were then checked (the inset of Figure 3c). In comparison to that of as grown CNFs films shown in Fig-ure 1b, no obvious surface damages or differences regarding surface morphology and porosity are observed, demonstrating the excellent stability of the CNFs/BDD capacitor electrode.

For the cycling stability test of a CNFs/BDD PC device, a long time charge/discharge cycling using GCD technique at the current density of 10 mA cm$^{-2}$ was carried out for 10000 cycles. The capacitance retention as a function of the cycle number is shown in Figure 4c, illustrating that the initial capacitance maintains unchanged after lifetime test. The surface morphologies of the two electrodes (the inset of Figure 4c) show almost no distinction, compared to the as grown CNFs films (Figure 1b), demonstrating the perfect stability of the electrodes and also a high degree of the reversibility in the electrolyte.

Therefore, one can conclude that both fabricated CNFs/BDD EDLC and PC devices have long-term cycle ability, one of important and necessary advantages of battery-like supercapacitors.

**Energy and Power Densities**

To clarify the overall performance of the CNFs/BDD EC devices, their energy and power densities were further calculated. The related Ragone plots are displayed in Figure 5. The estimated maximal *E* and *P* reach the values of 22.9 W h kg$^{-1}$ and 27.3 kW kg$^{-1}$ for an EDLC device, respectively, while for a PC device, they are 44.1 W h kg$^{-1}$ and 25.3 kW kg$^{-1}$, respectively. Compared to other energy devices (e.g., ECs, batteries, etc.) shown in Figure 5, the proposed CNFs/BDD EC devices exhibit not only much higher *P*, but also higher *E* than those for some of reported supercapacitors. The value of *E* is similar to that of batteries. The reason is ascribed to the structure of the electrodes and also the "battery-like" behavior of the $Fe(CN)_6^{3-/4-}$ redox electrolyte.[10a]

Table S1 further compares the performance (e.g., *E* and *P*) of our CNFs/BDD EC devices with that of other CNFs based EC devices reported in literature. The *E* and *P* values of our CNFs/BDD EC devices are higher than those of many reported CNFs based EC devices.[24] For instance, porous CNFs based EDLC devices show a highest *E* of 17 W h kg$^{-1}$ and a *P* of 20 kW kg$^{-1}$.[25] The *E* and *P* of EDLC devices fabricated by N,P – co-doped CNFs networks reach the values of 7.8 W h kg$^{-1}$ and 26.6 kW kg$^{-1}$, respectively.[26] By applying $V_2O_5$/CNFs composites as electrode material, the PC devices exhibit a high $E$ of 18.8 W h kg$^{-1}$, as well as a high $P$ of 20.0 kW kg$^{-1}$.[27]

In conclusion, both CNFs/BDD EDLC and PC devices exhibit high power densities together with high energy densities. They are characteristic for battery-like supercapacitors. Together with obtained large capacitances, long-term capacitance retention, battery-like EDLCs and PCs are successfully formed using vertically aligned carbon nanofibers grown on BDD.

**Battery-like Supercapacitor Demonstrator**

A stand-alone and portable system was designed to demonstrate the proposed CNFs/BDD ECs for practical application. Figure 6a shows the formed demonstrator, which consists of three EDLC devices assembled in series, a single-board microcontroller to control the charge/discharge process, a red LED (1.8 V), and a USB cable to charge this device. Figure 6b illustrates schematically the designed prototype of CNFs/BDD EC devices. The two CNFs/BDD electrodes were attached tightly to the both side of the cell, made from transparent acrylic glass. The efficient area of each electrode exposed to the electrolyte was about 0.785 cm$^2$. A 50 μm Nafion membrane was fixed with two sheets in the middle of the cell. As a case study, the electrolyte (1.0 M $H_2SO_4$) was filled in the cell from the top.

Figure 6c shows the circuit diagram of such an EC. In the first step, the switch '1' is closed, the device is charged by an external power supply with a USB connector. The resistance in the circuit is used to adjust the charging current, or charging duration. When the measured potential of the EC device is up to 3 V, the switch '1' opens and then the switch '2' is closed, leading to power a commercial red LED to illuminate. Moreover, the discharge process of the EC device is possible to be tested. When the potential is lower than 1.6 V, the switch '2' opens and the switch '1' is closed, and the EC devices are charged again. Such a process is automatically controlled by a single-board microcontroller which is also possible to be connected with a USB connector to a computer (Figure 6a). Figure 6d shows the variation of potentials in the recorded curves during the charge/discharge process as a function of the time. The good re-peatability of the curves reveals the excellent reversibility of the device. With a charge time of about 70 s, a red LED (working voltage: 1.8 V) lights up (Figure 6d). The light intensity is varied as a function of the applied voltages (the insets in Figure 6d). With a high voltage at an initial stage, it is very bright, hinting the features of an EC. The light lasts for few seconds and becomes weaker till it is fully out. These confirm the battery-like properties of the EC. Therefore, the fabricated battery-like supercapacitors from CNFs/BDD hybrid films are promising for the employed for practical energy storage applications.

## Conclusion

Vertically aligned CNFs have been directly grown on BDD and been further utilized as a novel hybrid carbon material to develop ECs. Both EDLCs using 1.0 M $H_2SO_4$ aqueous solution and PCs using 0.05 M $Fe(CN)_6^{3-/4-}$ + 1.0 M $Na_2SO_4$ have been fabricated. These ECs feature large capacitance (in the range of mF cm$^{-2}$), long-term capacitance retention, high power densities and high energy densities. For example, the assembled two-electrode symmetrical EC devices exhibit high specific capacitances of 30.0 and 48.0 mF cm$^{-2}$ at a scan rate of 10 mV s$^{-1}$ in the inert aqueous solution and the redox active electrolyte, respectively. The capacitances keep unchanged even after 10000 cycles for both EDCL and PC devices. For EDLC devices, the energy and power densities were 22.9 W h kg$^{-1}$ and 27.3 kW kg$^{-1}$, respectively. While for PC devices, they were 44.1 W h kg$^{-1}$ and 25.3 kW kg$^{-1}$, respectively. Such high energy and power densities achieved from the CNFs/BDD EC devices are comparable to those of batteries. Moreover, the performance of these battery-like supercapacitors is possible to be further enhanced only with longer CNFs (e.g., simply by applying a longer growth time). The high performance of these ECs has been intercepted by the large surface areas, improved electrical conductivity, and porous structures of these binder-free CNFs/BDD hybrid films. A stand-alone demonstrator further confirms the suitability of these battery-like supercapacitors for powering future multifunctional electronics, hybrid electric vehicles, and industrial equipment.

## Acknowledgement

S.Y. gratefully acknowledges the financial support from China Scholarship Council (Chinese Government Scholarship, Award no. 201408080015). N.Y. acknowledges the financial support from the German Research Foundation (DFG) under project YA344/1-1.

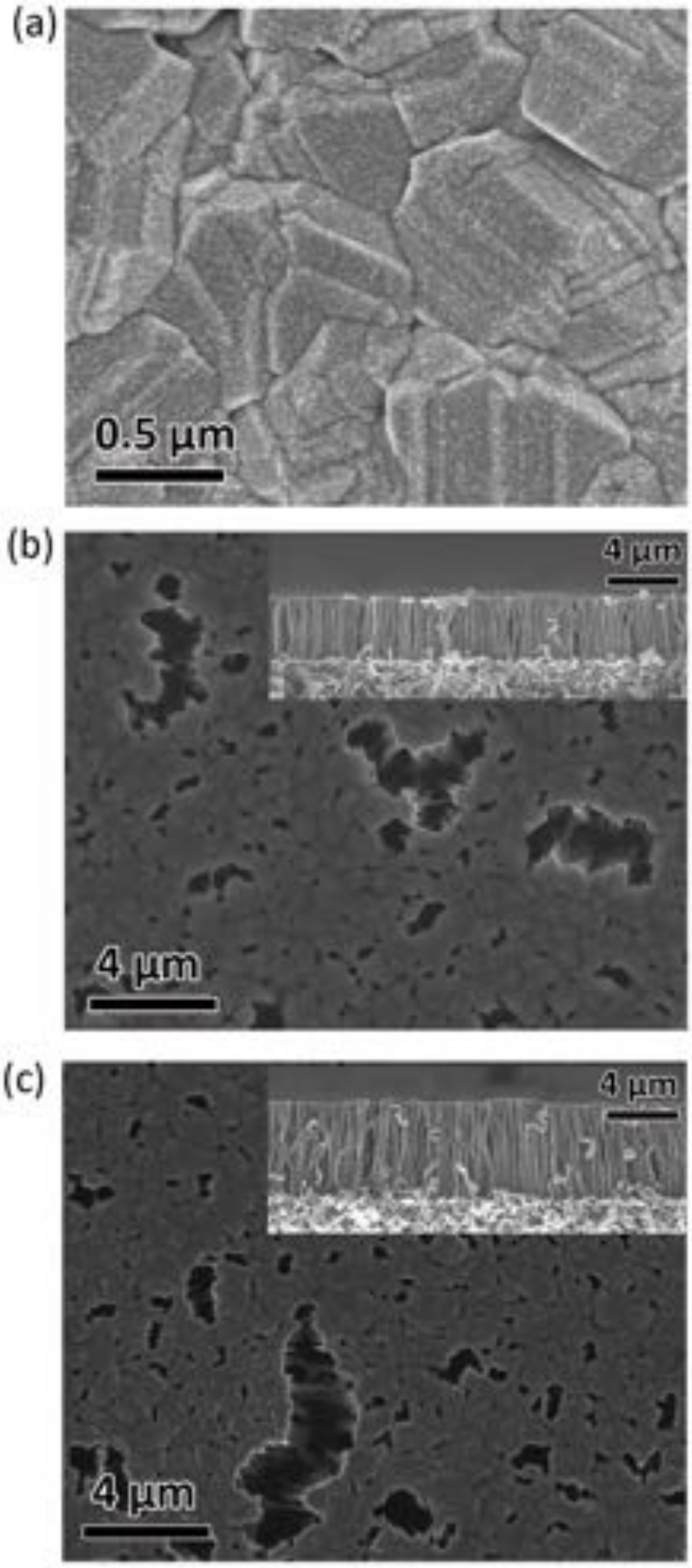

Figure 1. Characterization: SEM images of (a) Cu film on BDD, (b) CNFs film on BDD with a growth time of 60 min, (c) CNFs film with a growth time of 90 min. The time for Cu sputtering was 60 s. The inset images show the cross sections of as-grown CNFs.

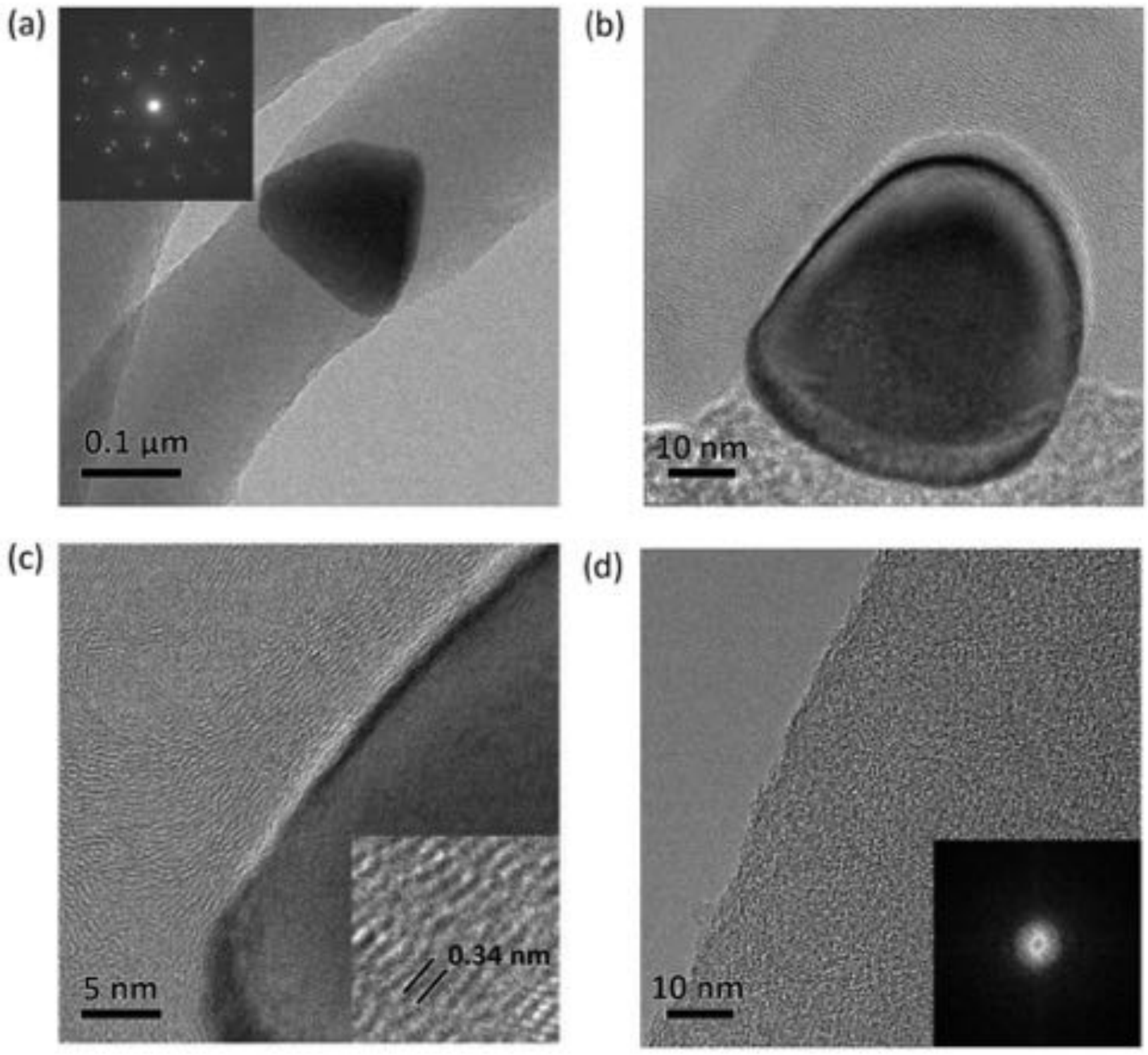

Figure 2. Characterization: (a) TEM image of one CNF with a Cu catalyst. The inset shows the SAED pattern of a Cu catalyst. (b) TEM image of the interface between one CNF and a Cu catalyst. (c) High-magnification TEM image of image b. The inset reveals the lattice fringes of graphite. (e) HRTEM and FFT (inset) images of the amorphous phase in the CNF.

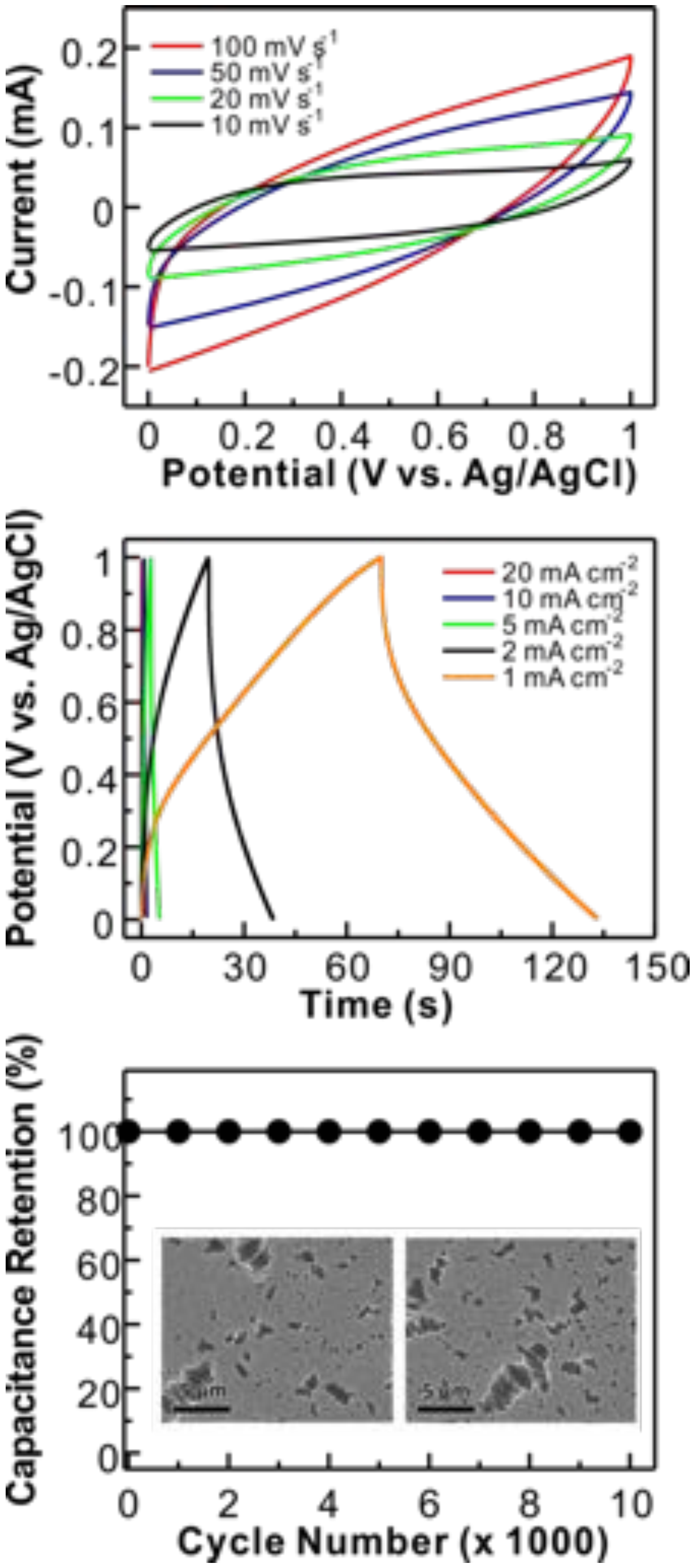

Figure 3. Performance of a CNFs/BDD symmetric EDLC device in 1.0 M $H_2SO_4$: (a) CVs recorded at the scan rates of 100, 50, 20, and 10 mV $s^{-1}$; (b) Charge/discharge curves at the current densities of 1, 2, 5, 10, and 20 mA $cm^{-2}$; (c) Capacitance retention at the charge/discharge current density of 5 mA $cm^{-2}$. The inset SEM images show the morphologies of the two used CNFs/BDD hybrid films after 10 000 charge/discharge cycles.

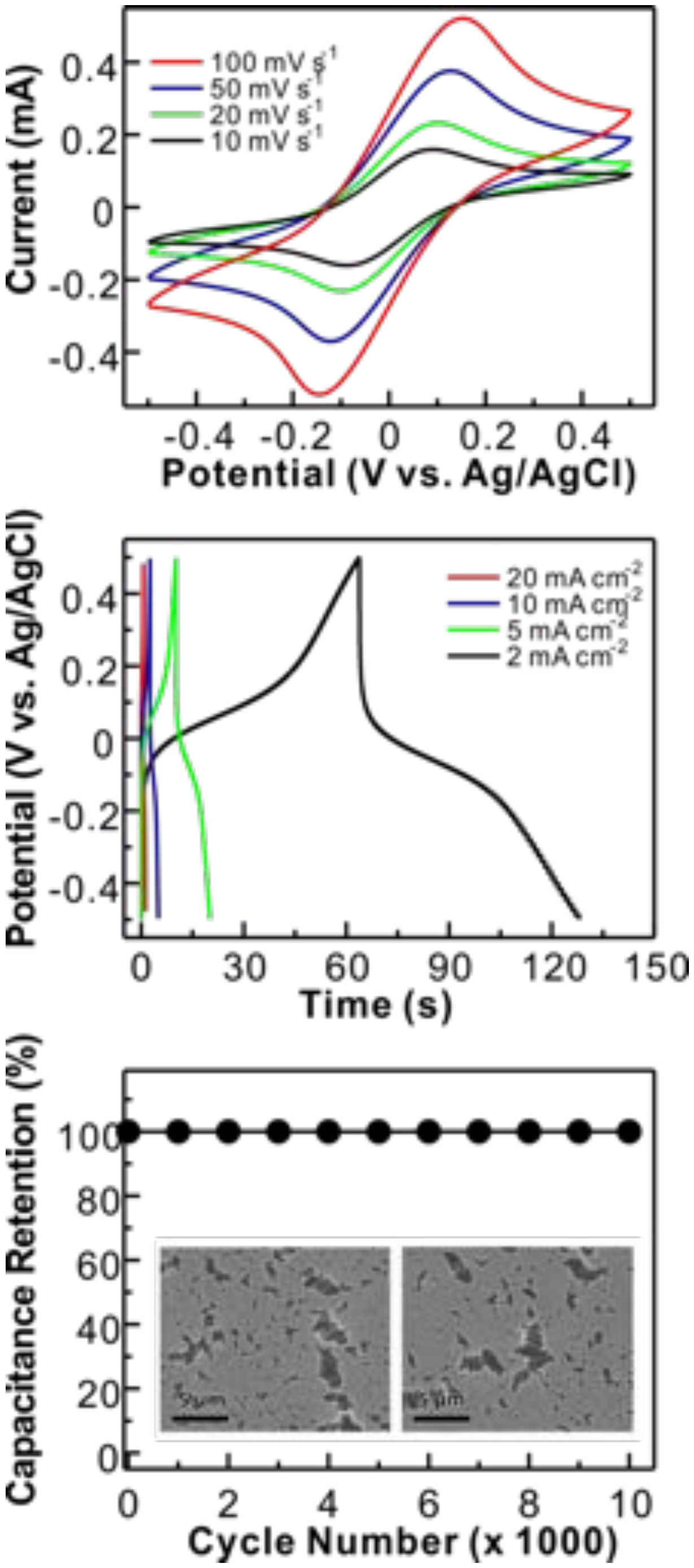

Figure 4. Performance of a CNFs/BDD symmetric PC device in 0.05 M $Fe(CN)_6^{3-/4-}$ + 1.0 M $Na_2SO_4$: (a) CVs at the scan rates of 100, 50, 20, and 10 mV $s^{-1}$; (b) Charge/discharge curves at the current densities of 2, 5, 10, and 20 mA $cm^{-2}$; (c) Capacitance retention at the charge/discharge current density of 10 mA $cm^{-2}$. The inset SEM images show the surface characteristics of the two used CNFs/BDD hybrid films after 10 000 charge/discharge cycles.

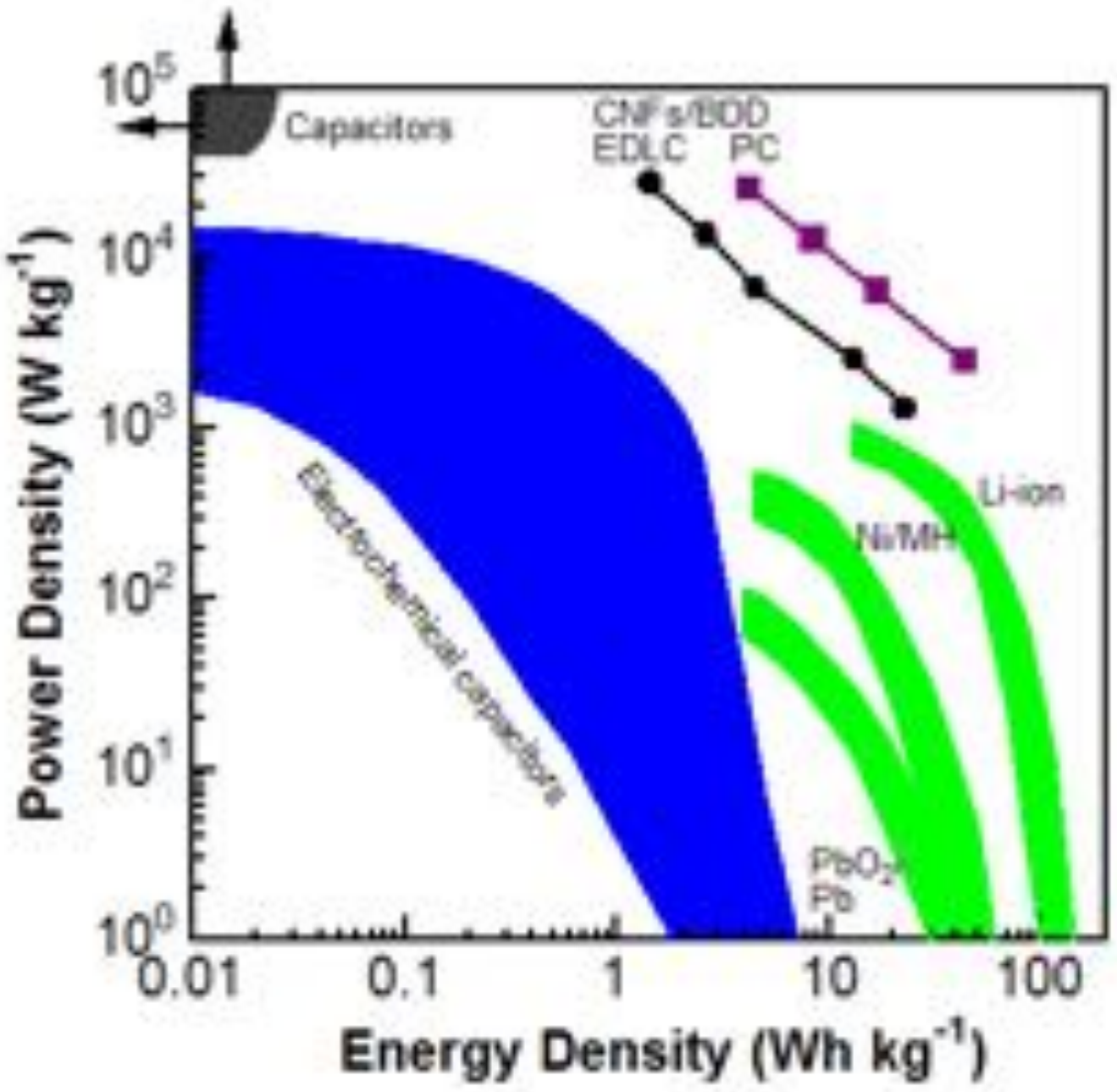

Figure 5. Ragone plots of CNFs/BDD EDLCs and PCs, traditional capacitors, other ECs and batteries, etc.[28]

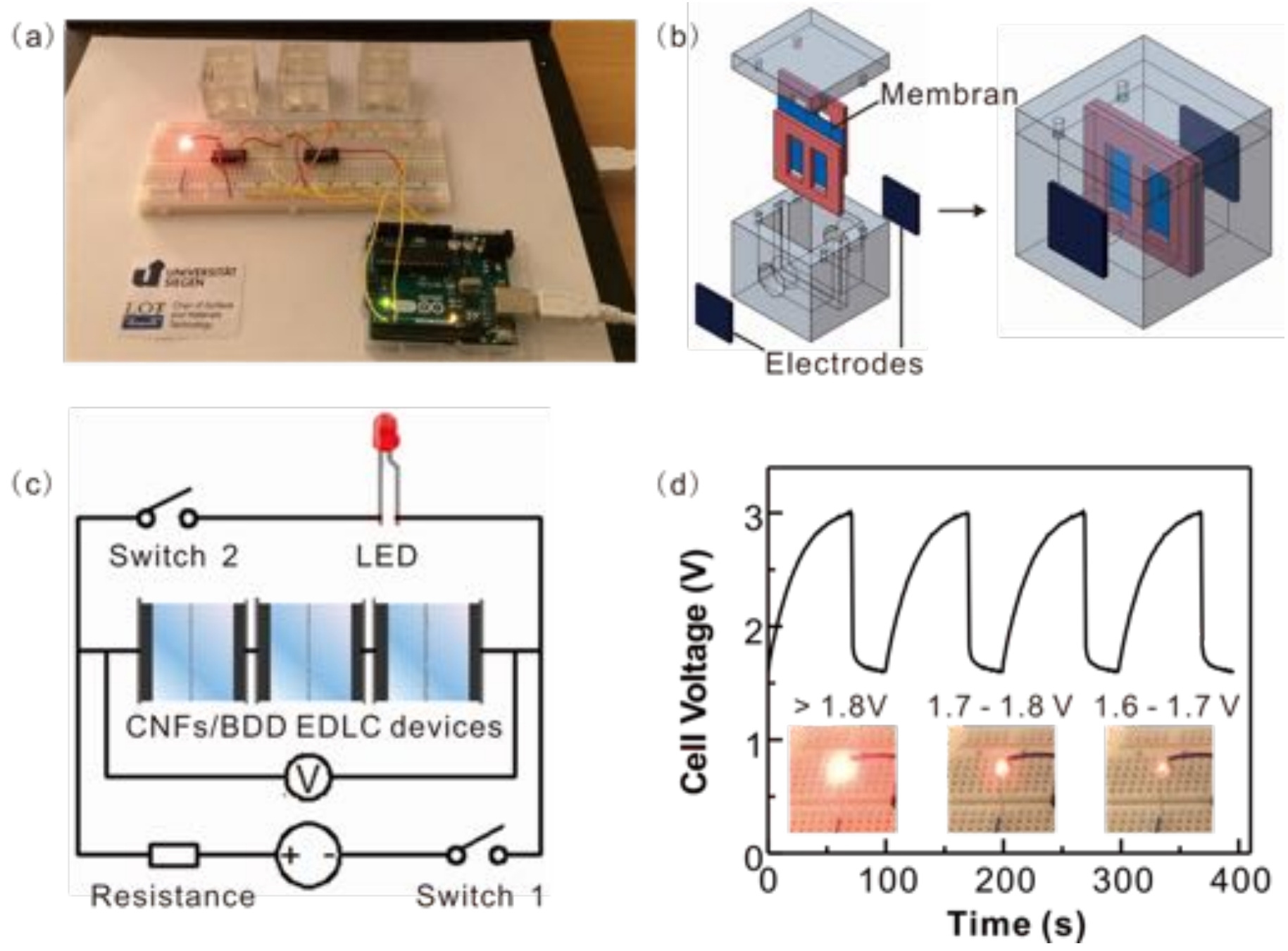

Figure 6. A stand-alone CNFs/BDD EC demonstrator: (a) a photograph of the demonstrator consisting of three CNFs/BDD EDLC devices in series, a single-board microcontroller connected to the computer with a USB cable, and a red LED; (b) design of a CNFs/BDD EC prototype used in the system; (c) schematic electric circuit diagram related to (a); (d) typical curves of the potential as a function of time during the charge/discharge processes. The insets show the variation of light intensity of the red LED in related to the cell voltage.

Supplementary Information

# Battery-like Supercapacitors from Vertically Aligned Carbon Nanofibers Coated Diamond: Design and Demonstrator

Siyu Yu,[a] Nianjun Yang,[a*] Michael Vogel,[a] Soumen Mandal,[b] Oliver A. Williams,[b] Siyu Jiang,[c] Holger Schönherr,[c] Bing Yang,[d] and Xin Jiang[a,d*]

[a] Institute of Materials Engineering, University of Siegen, 57076 Siegen, Germany

[b] School of Physics and Astronomy, Cardiff University, Cardiff CF24 3AA, UK

[c] Physical Chemistry I, Department of Chemistry and Biology, University of Siegen, 57076 Siegen, Germany

[d] Shenyang National Laboratory for Materials Science, Institute of Metal Research (IMR), Chinese Academy of Sciences (CAS), No.72 Wenhua Road, Shenyang 110016 China

E-mail: nianjun.yang@uni-siegen.de, xin.jiang@uni-siegen.de

**Experimental**

**Figures**

**Tables**

**Supporting references**

**Experimental**

The growth of CNFs using Cu particles as the catalysts was performed. Cu particles were simply obtained through thermal annealing of the sputter copper films under the conditions of an annealing temperature of 500 °C, a pressure of about $5\times10^{-2}$ mbar, and an annealing time for 60 min. SEM images of these annealed Cu films are shown in Figure S5. During the thermal treatment, Cu films split into fragments. Some of fragments further aggregate into separate particles. For Cu films with $t_{Cu,s}$ from 15 to 60 s, the converged Cu particles feature larger sizes and higher densities, which are increased as a function of $t_{Cu,s}$. For Cu films with $t_{Cu,s}$ from 90 to 120 s, the disintegrated fragments are not completely converted into drop-shaped elements. This is due to the increased thicknesses of copper films and insufficient annealing times.

SEM images of CNFs grown on these annealed Cu films are presented in Figure S6 (for $t_{Cu,s}$ from 15, 30, 60, 90, and 120 s). The inset images show the cross sections of the related films. Vertically aligned CNFs are gained with a $t_{Cu,s}$ of 90 and 120 s. This is attributed to the nearly continuous dispersion of Cu fragments, resulting in analogous growth phenomenon as that without annealing of the Cu films (WOA). For CNFs/BDD hybrid films with $t_{Cu,s}$ from 15 to 60 s, disordered CNFs with different dimensions are obtained, closely associated with the distribution and size of Cu particles. The thicknesses of CNFs films are about 2.2, 2.5, 3.0, 3.2, and 3.3 μm for $t_{Cu,s}$ of 15, 30, 60, 90, and 120s, respectively. These values are slightly smaller than those CNFs films WOA. The difference is originated from the influence of varied sizes of Cu catalysts on the growth rates of the CNFs.[1]

**Figures**

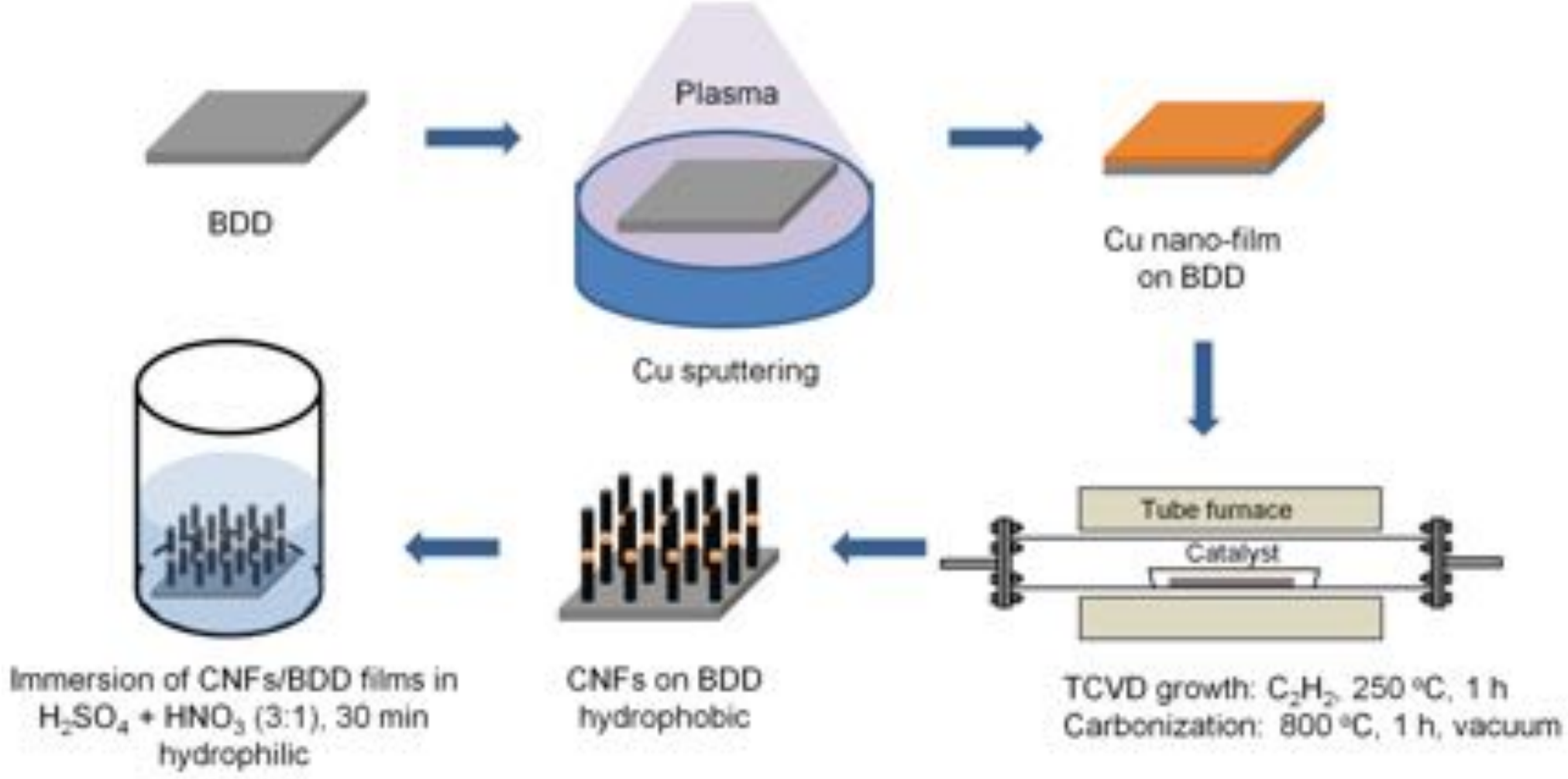

Figure S1. Schematic plots of the synthesis of the CNFs/BDD hybrid films.

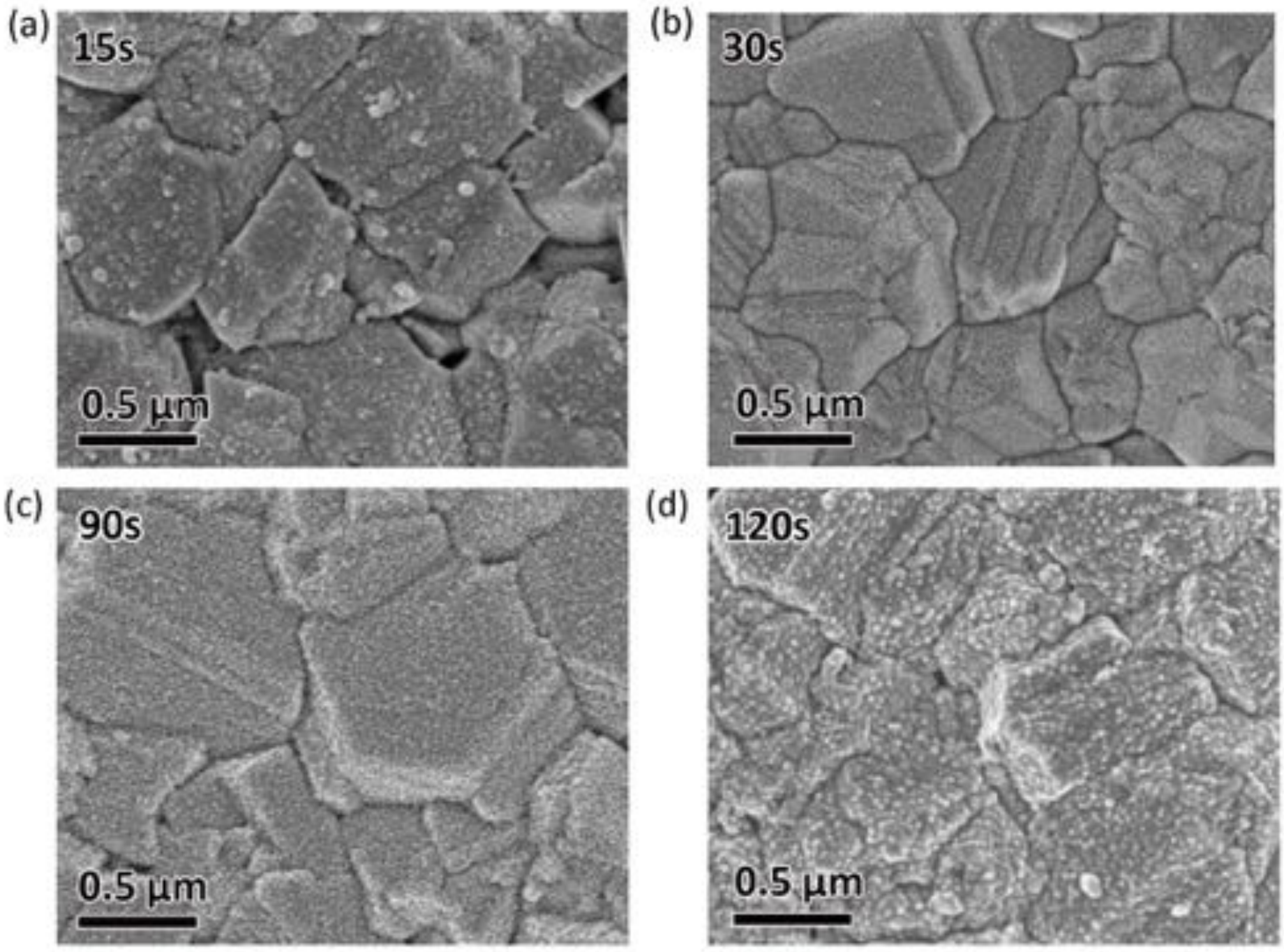

Figure S2. SEM images of Cu films on BDD with a sputtering time ($t_{Cu,s}$) of (a) 15, (b) 30, (c) 90, and (d) 120 s.

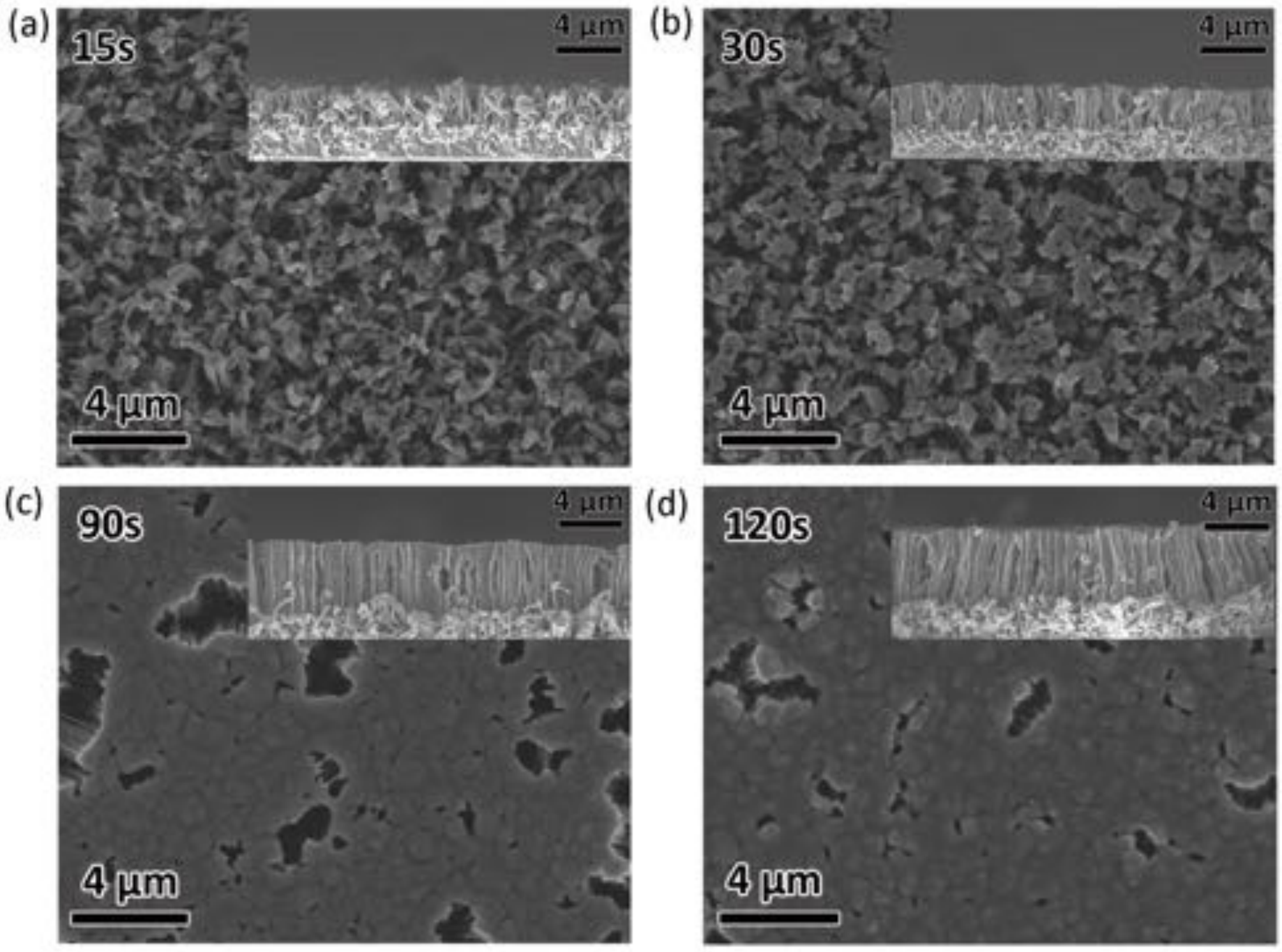

Figure S3. SEM images of the CNFs/BDD hybrid films with $t_{Cu,s}$ of (a) 15, (b) 30, (c) 90, and (d) 120 s. The insets are their side-view SEM images.

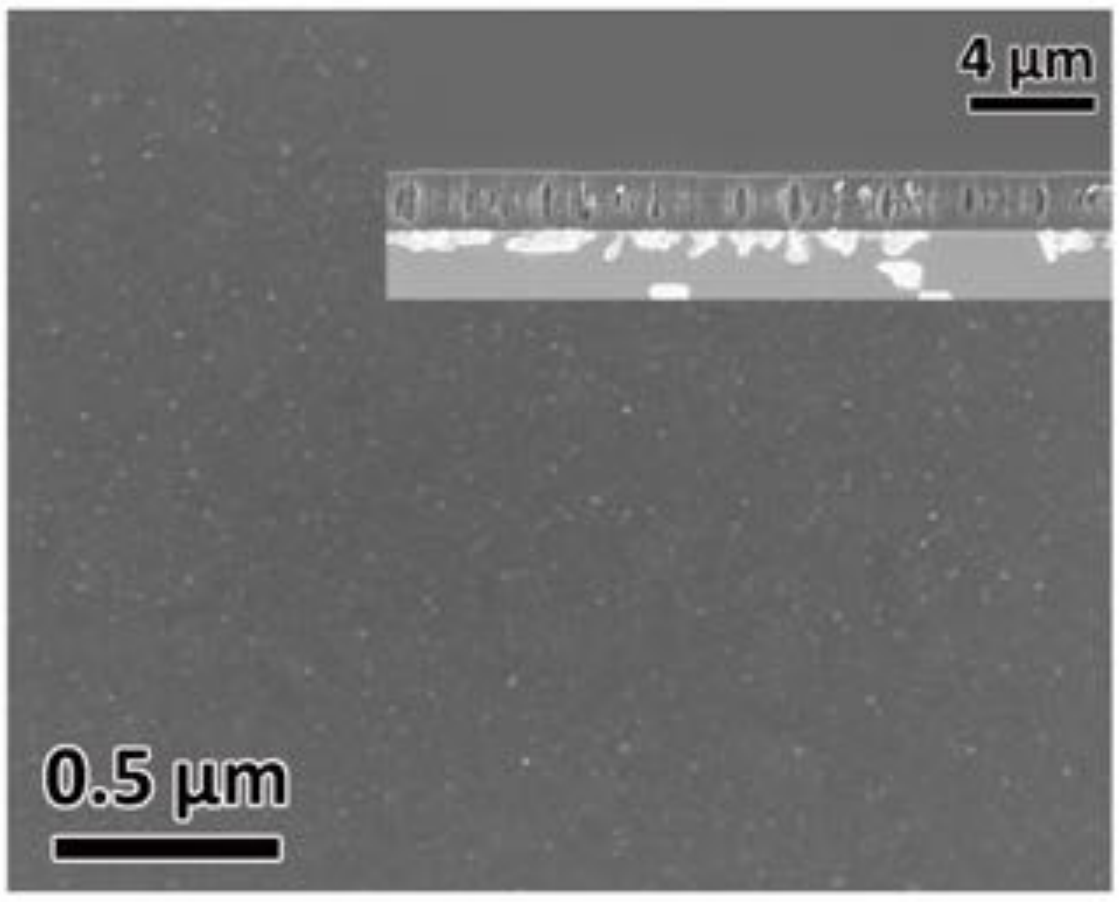

Figure S4. SEM image of a CNFs film grown on a smooth Si substrate with a $t_{Cu,s}$ of 60 s, WOA. The inset shows its cross section.

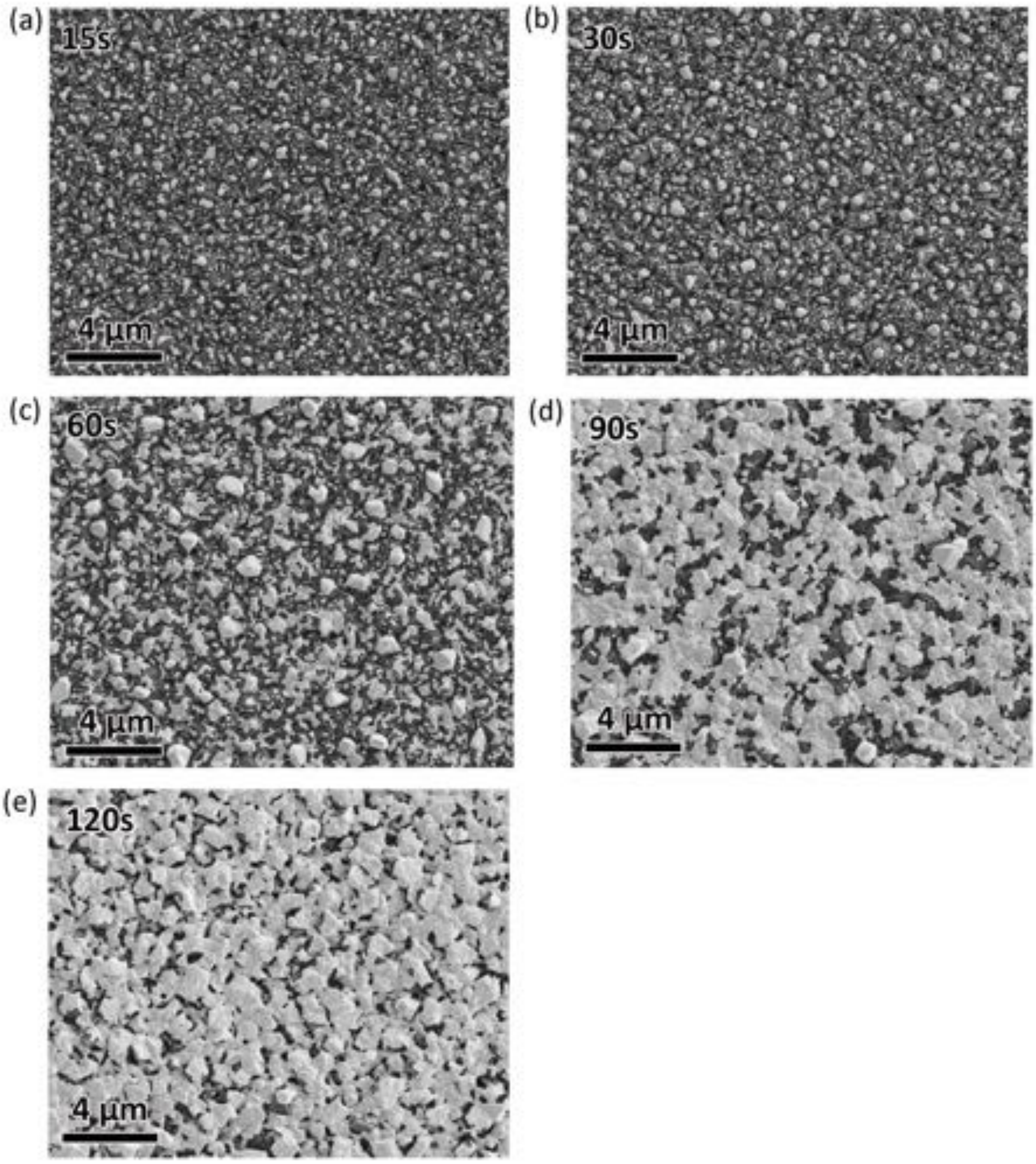

Figure S5. SEM images of Cu films with a $t_{Cu,s}$ of (a) 15, (b) 30, (c) 60, (d) 90, and (e) 120 s after annealing at 500 $^{o}$C for 1 h in vacuum.

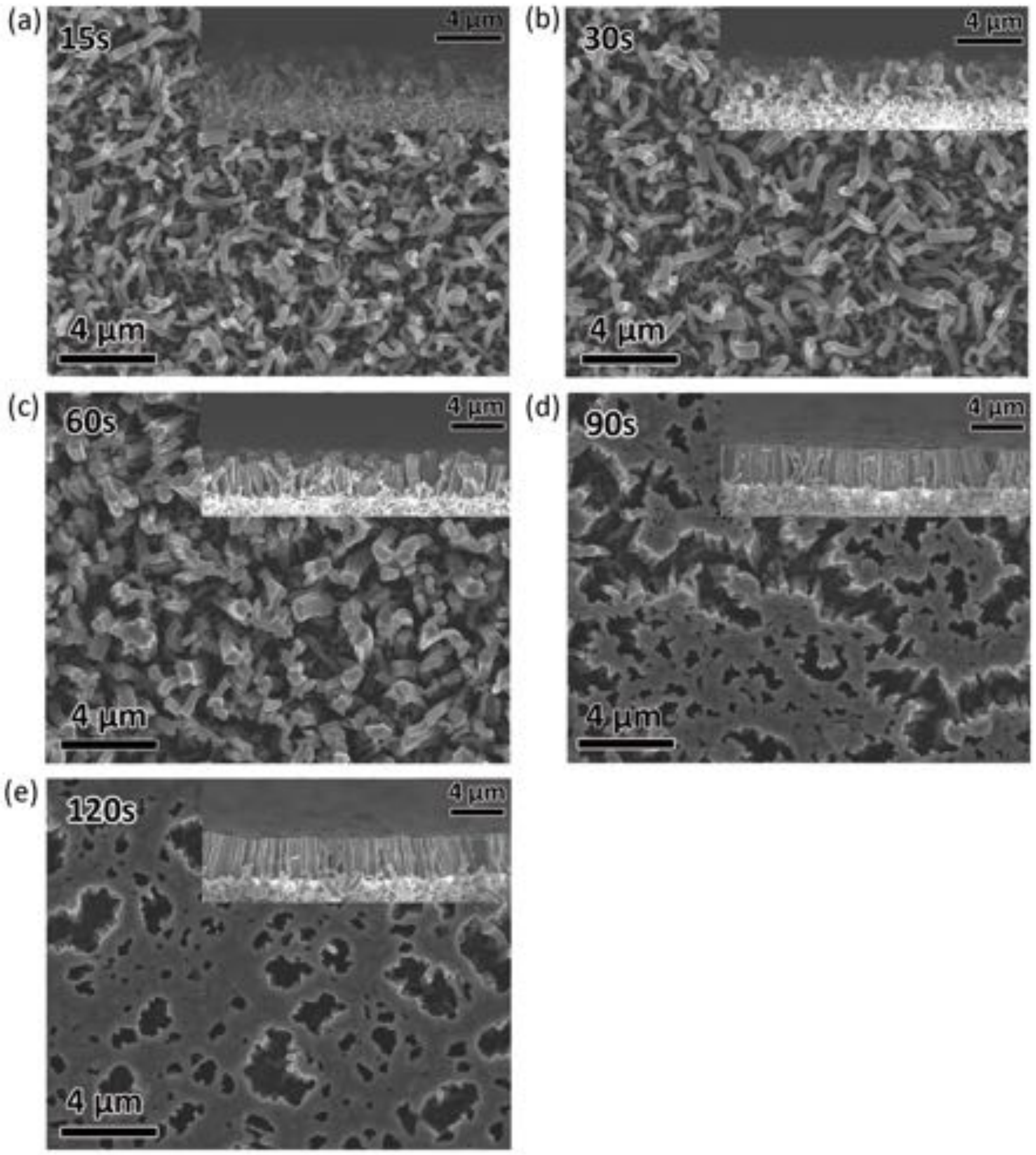

Figure S6. SEM images of the CNFs/BDD hybrid films grown after the annealing of Cu films with a $t_{Cu,s}$ of (a) 15, (b) 30, (c) 60, (d) 90, and (e) 120 s. The insets show the cross sections.

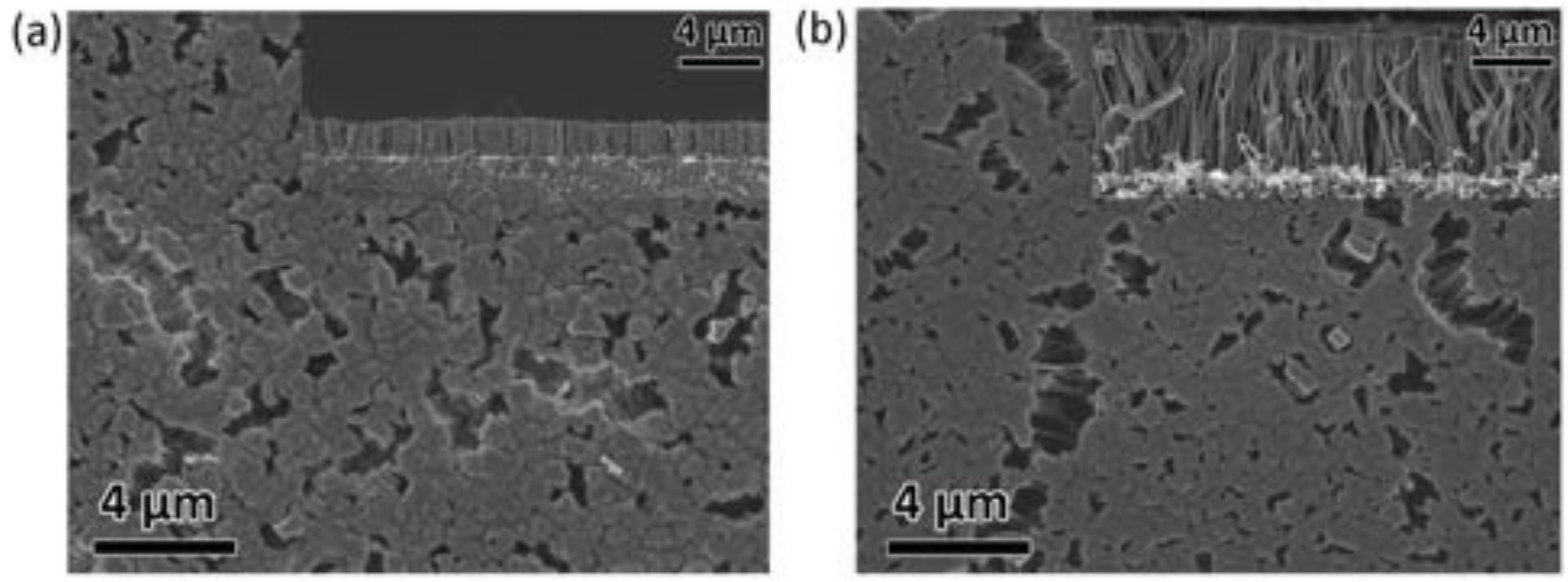

Figure S7. SEM images of the CNFs/BDD hybrid films with a $t_{Cu,s}$ of 60 s WOA with the CNFs growth time of (a) 30 and (b) 120 min.

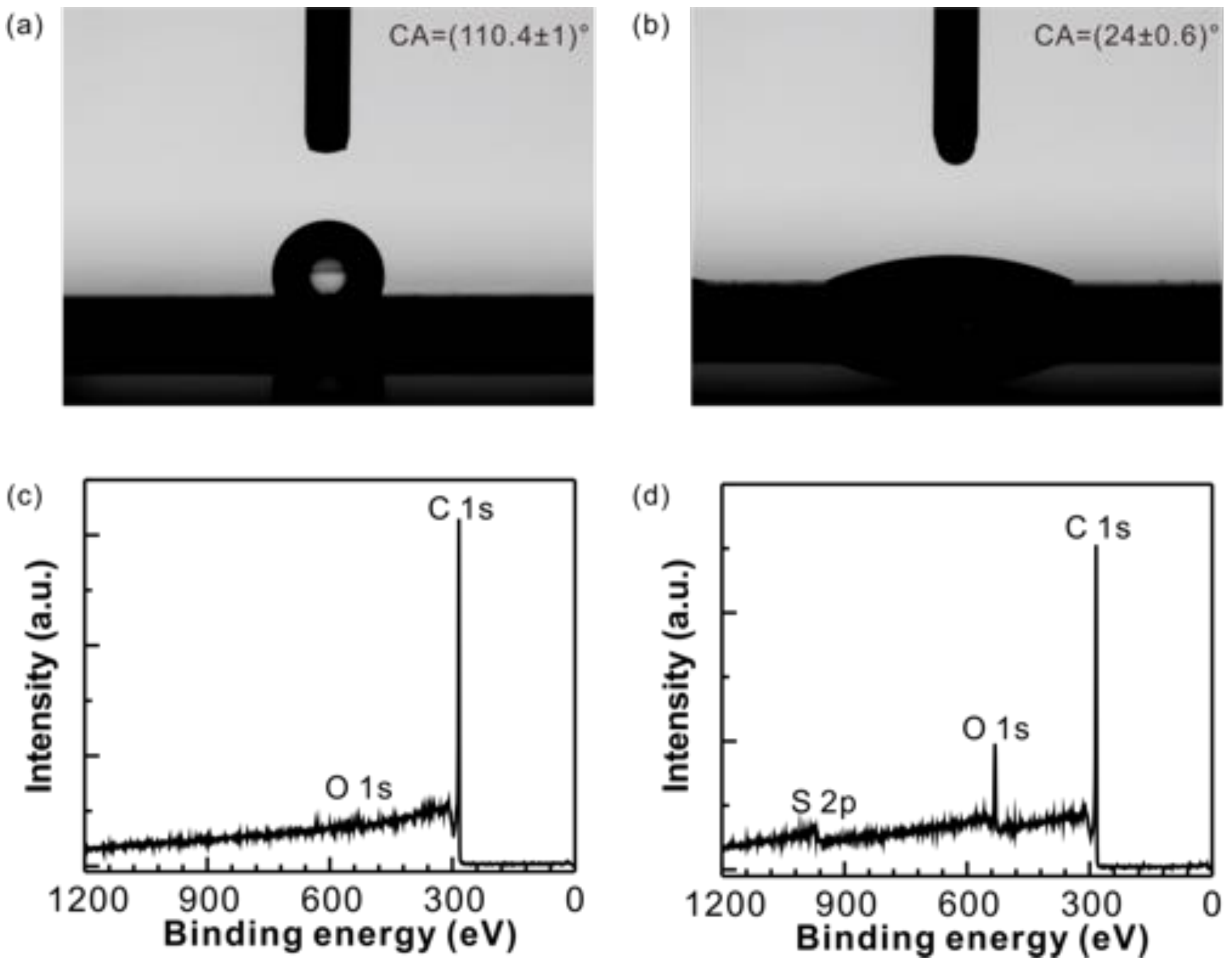

Figure S8. (a,b) Contact angels and (c.d) XPS of the CNFs/BDD hybrid films with a $t_{Cu,s}$ of 60 s WOA: (a,c) as grown, (b,d) after treatment in sulfuric and nitric acid (v:v = 3:1) for 30 min.

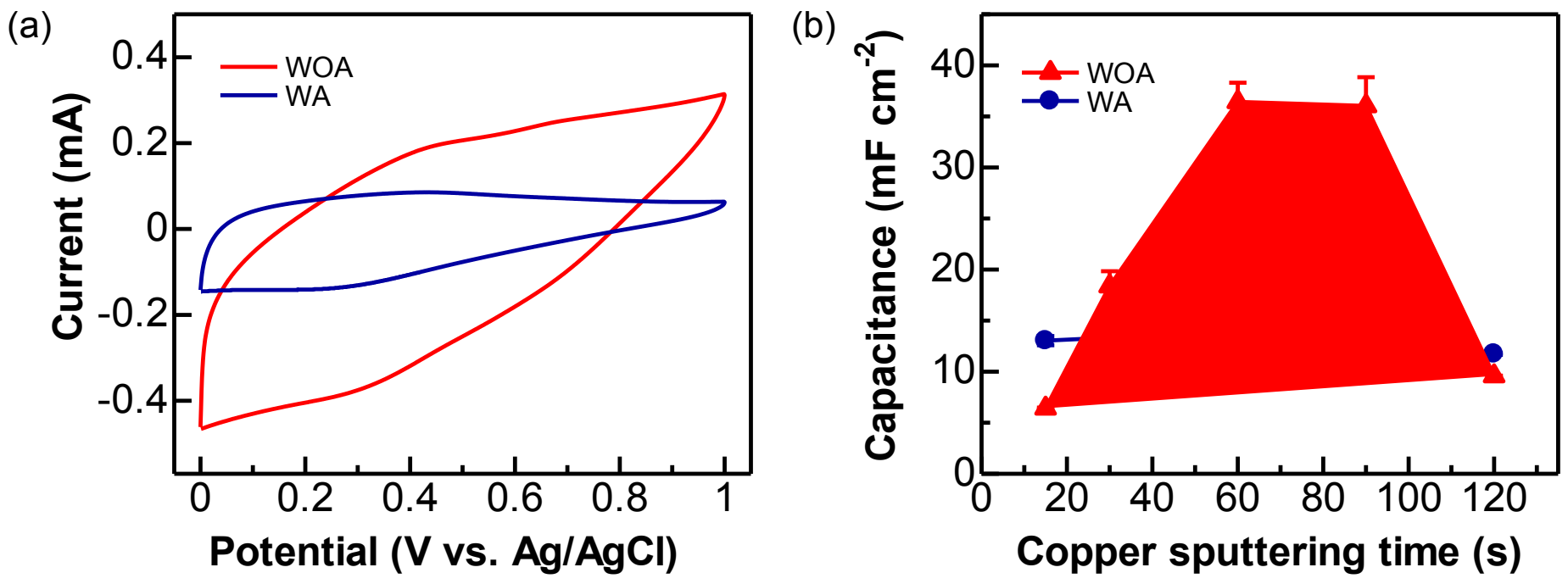

Figure S9. (a) Cyclic voltammogramms of the CNFs/BDD hybrid films WA and WOA with a $t_{Cu,s}$ of 60 s at a scan rate of 100 mV $s^{-1}$ in 1.0 M $H_2SO_4$ aqueous solution. (b) Comparison of the capacitances of the CNFs/BDD hybrid films with (WA) and without (WOA) annealing of Cu films with different $t_{Cu,s}$.

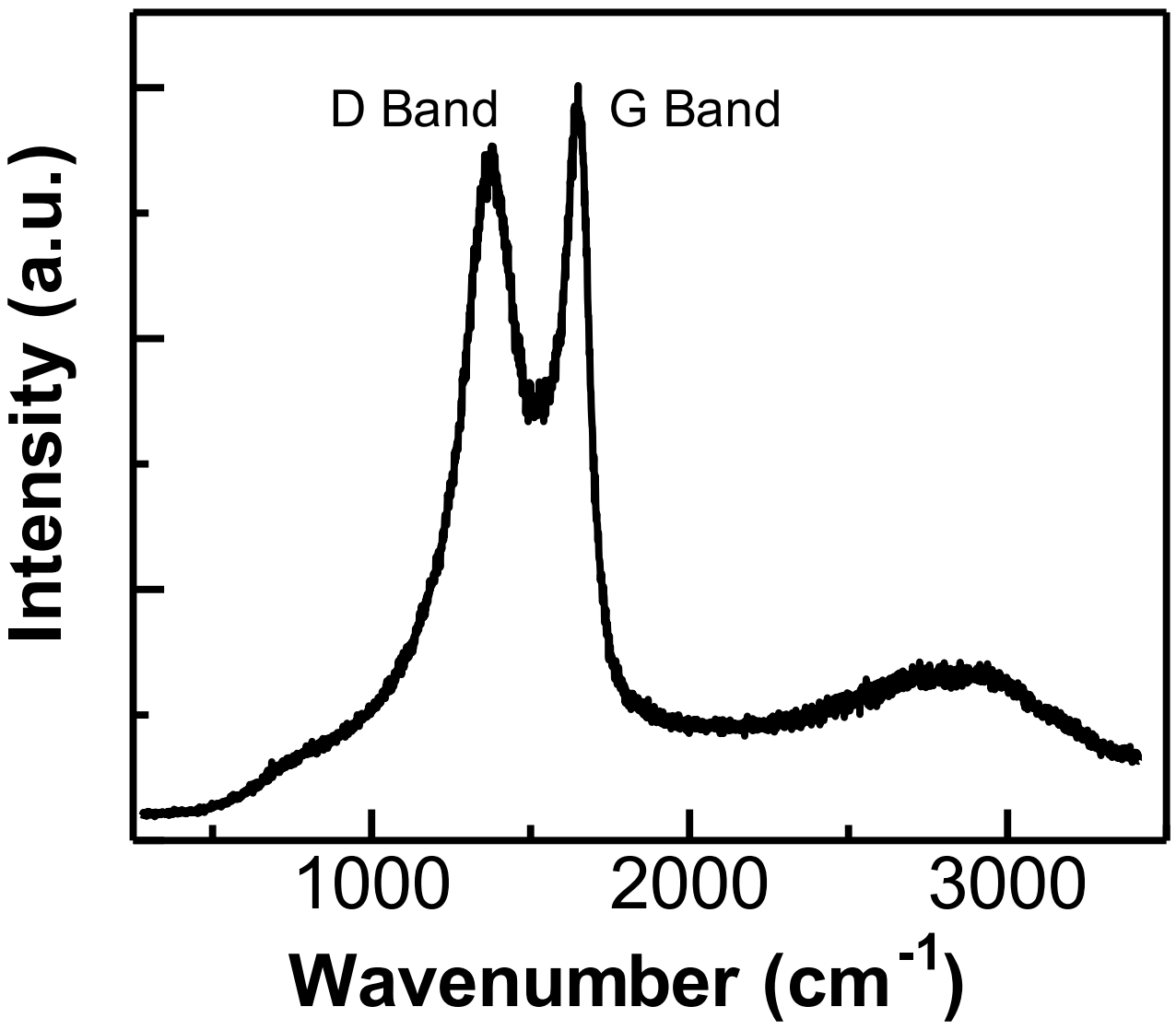

Figure S10. The Raman spectrum of a CNFs/BDD hybrid film with a $t_{Cu,s}$ of 60 s WOA.

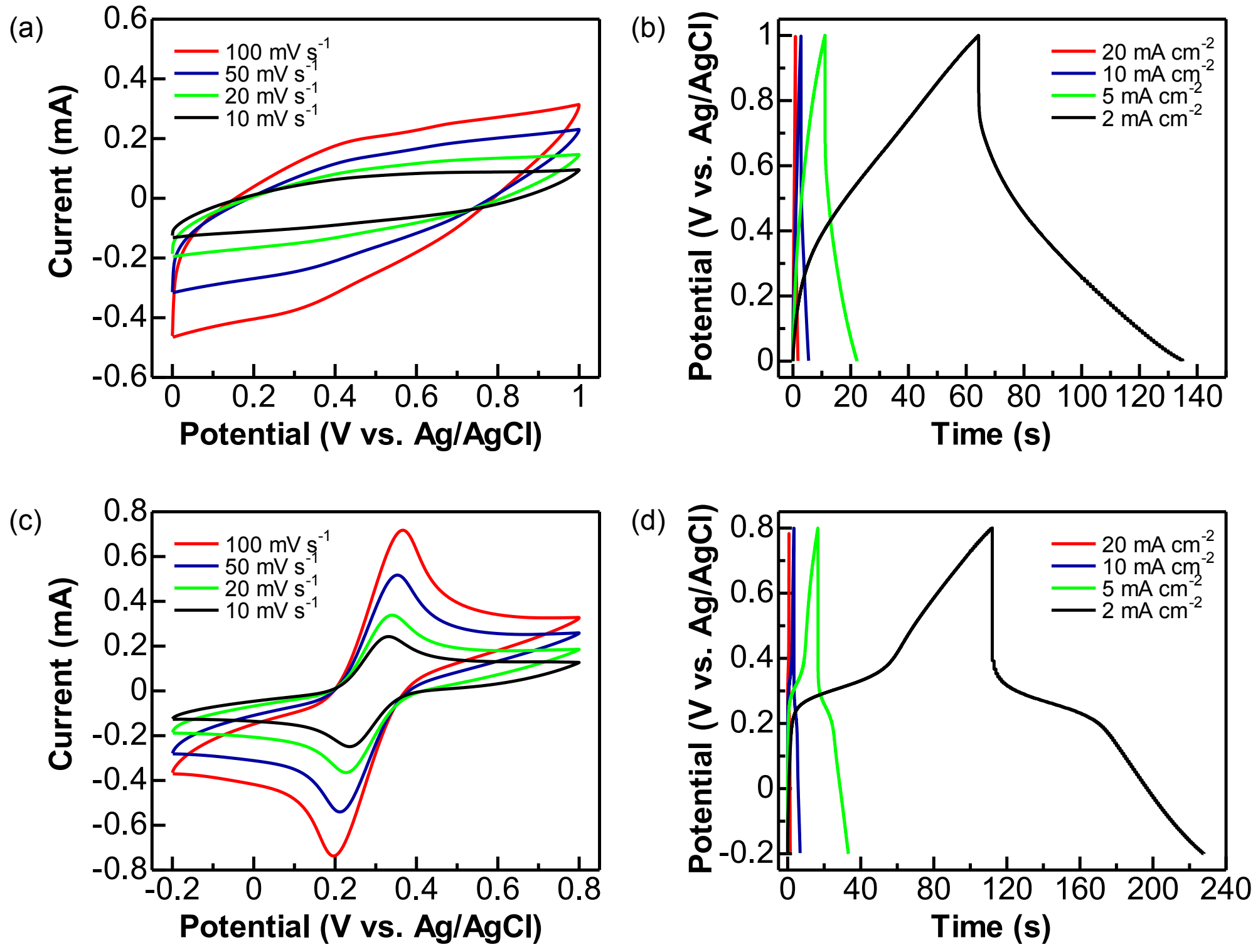

Figure S11. Performance of CNFs/BDD electrochemical capacitors (ECs) with a $t_{Cu,s}$ of 60 s WOA using a three-electrode system in (a, b) 1.0 M $H_2SO_4$ inert and (c, d) 0.05 M $Fe(CN)_6^{3-/4-}$ + 1.0 M $Na_2SO_4$ redox electrolytes. (a, c) Cyclic voltammograms at the scan rates from 10 to 100 mV $s^{-1}$. (b, d) Charge-discharge curves at the current densities from 1 to 20 mA $cm^{-2}$.

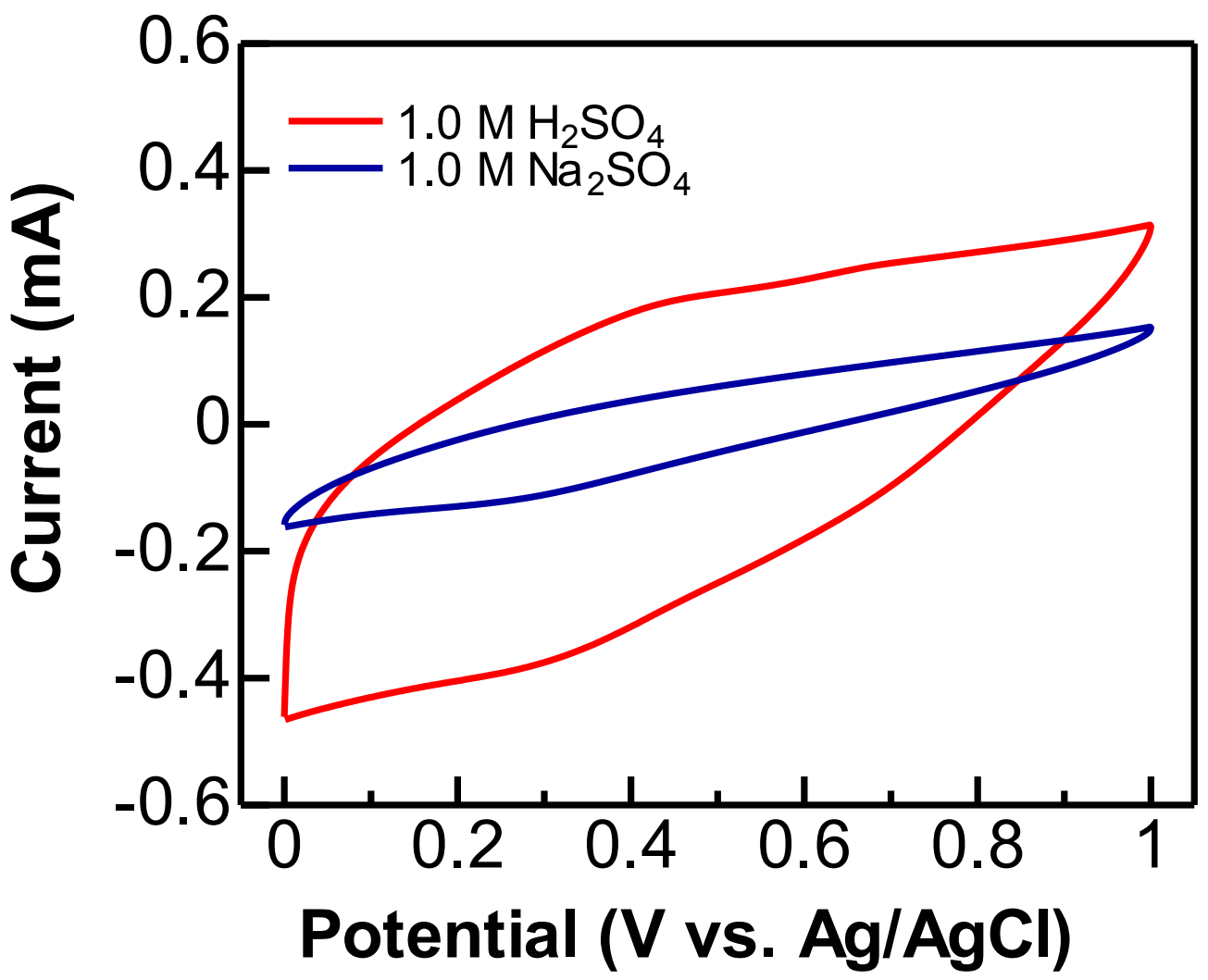

Figure S12. Cyclic voltammograms of a CNFs/BDD hybrid film with a $t_{Cu,s}$ of 60 s WOA in the electrolyte of 1.0 M $Na_2SO_4$ and 1.0 M $H_2SO_4$ at a scan rate of 100 mV $s^{-1}$.

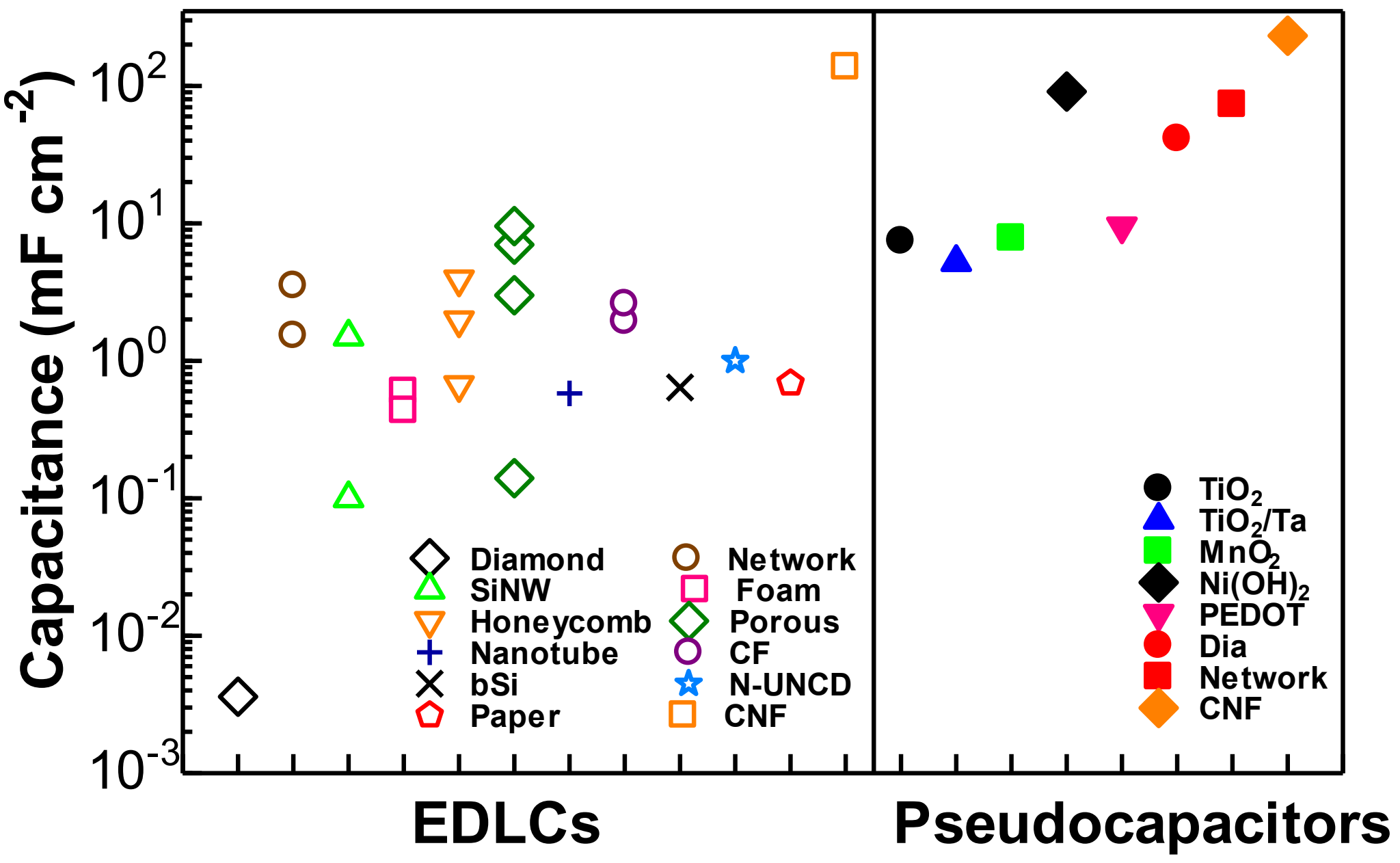

Figure S13. Comparison of the capacitances of diamond nanostructures based electric double layer capacitors (EDLCs) and pseudocapacitors (PCs). The utilized nanostructures are: diamond[2], diamond network (network)[3], diamond/silicon nanowires (SiNW)[4], diamond foam (Foam)[5], honeycomb diamond (Honeycomb)[6], porous diamond (Porous)[7], BDD/Nanotube (Nanotube)[8], BDD/carbon fiber (CF)[9], BDD/'black silicon' (bSi)[10], nitrogen included ultrananocrystalline diamond (N-UNCD)[11], diamond paper (Paper)[12], CNFs/BDD (CNF, this work), BDD/TiO$_2$ (TiO$_2$)[13], TiO$_2$/BDD/Ta (TiO$_2$/Ta)[14], MnO$_2$/BDD (MnO$_2$)[2], Ni(OH)$_2$/Diamond Nanowire (Ni(OH)$_2$)[15], poly(3,4-(ethylenedioxy)thiophene)-coated diamond@silicon nanowires (PEDOT)[16]; BDD based PC using redox electrolyte (Dia)[3a], diamond network PC using redox electrolyte (Network)[3a], CNFs/BDD PC using redox electrolyte (CNF, this work).

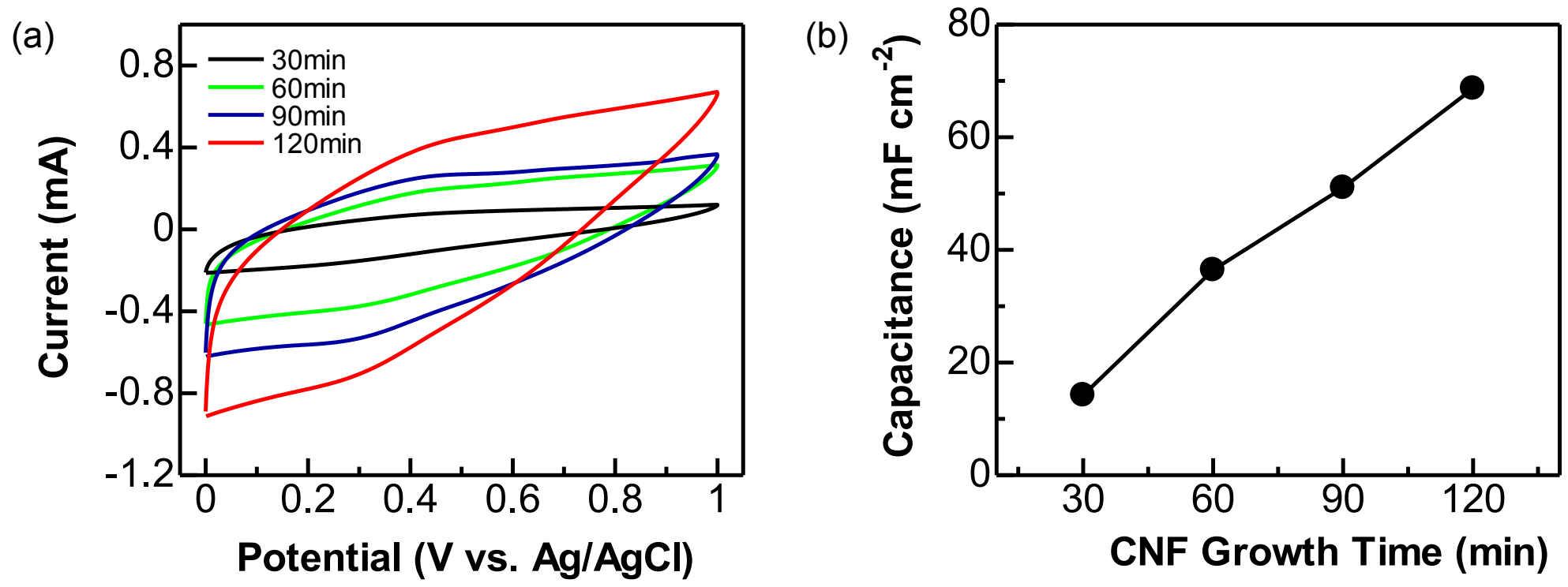

Figure S14. (a) Cyclic voltammograms of the CNFs/BDD hybrid films with a $t_{Cu,s}$ of 60 s WOA with the CNFs growth time of 30, 60, 90, and 120 min, The scan rate is 100 mV s$^{-1}$ and a three-electrode system is used. (b) Calculated capacitances from the related CVs in (a).

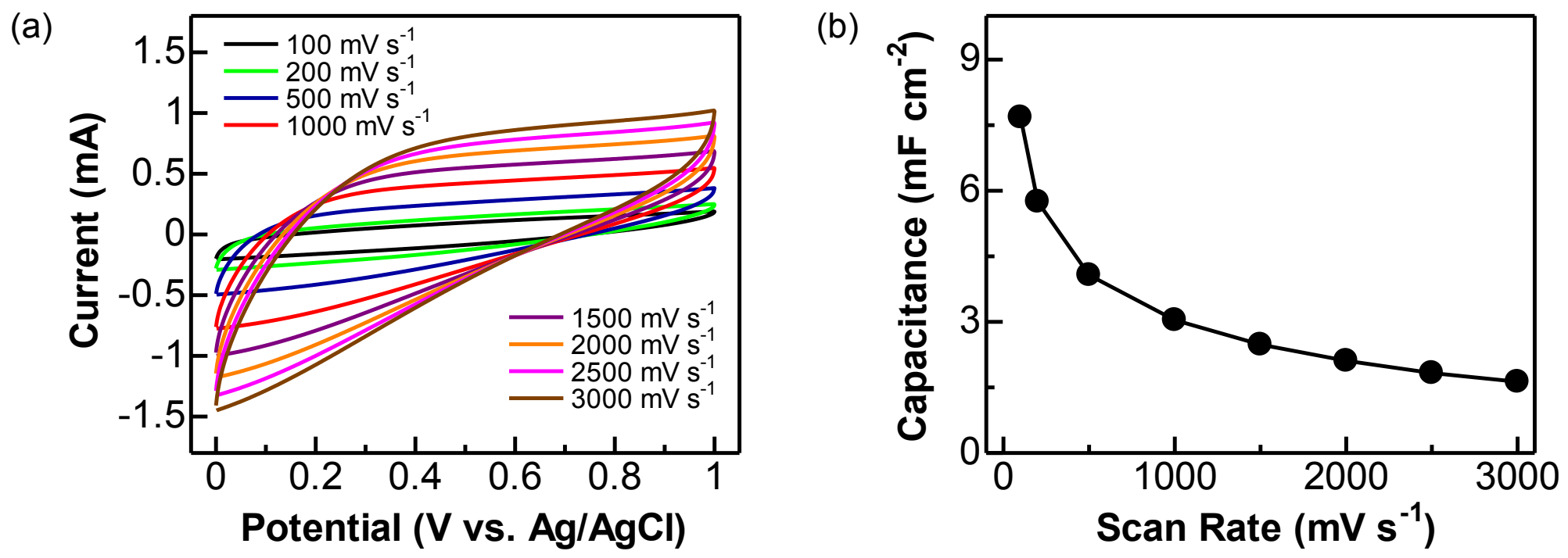

Figure S15. (a) Cyclic voltammograms of a CNFs/BDD EDLC device at the scan rates from 100 to 3000 mV s$^{-1}$ in 1.0 M $H_2SO_4$ solution. (b) Calculated capacitances from the related CVs in (a).

## Tables

Table S1. Comparison of the electrochemical performance of CNFs/BDD EC devices with that of other CNFs based EC devices

| | Electrode Material | Electrolyte | *E* / W h $kg^{-1}$ | *P* / kW $kg^{-1}$ | Ref-erence |
|---|---|---|---|---|---|
| EDLC | Hollow particle based N-doped CNF | 2.0M $H_2SO_4$ | 11.0 | 25.0 | [17] |
| | N,P - co-doped CNF networks | 2.0 M $H_2SO_4$ | 7.8 | 26.6 | [18] |
| | Cross-linked N-doped CNF network | 1.0M $H_2SO_4$ | 5.9 | 10.0 | [19] |
| | Mesoporous CNF | 6.0 M KOH | 5.1 | 16.0 | [20] |
| | CNF paper | 6.0 M KOH | 7.1 | 9.0 | [21] |
| | Porous CNF composites | 6.0 M KOH | 17.0 | 20.0 | [22] |
| | Porous CNF | 1.0 M $H_2SO_4$ | ~6.0 | 20.0 | [23] |
| | CNFs/BDD | 1.0M $H_2SO_4$ | **22.9** | **27.3** | This work |
| PC | $V_2O_5$/CNF composites | 6.0 M KOH | 18.8 | 20.0 | [24] |
| | ZnO/porous activated CNF | 6.0 M KOH | 22.7 | 4.0 | [25] |
| | $MnO_2$/CNF // CNF | 0.5 M $Na_2SO_4$ | 36 | 4.4 | [26] |
| | CNFs/BDD | 1.0M $Na_2SO_4$ + 0.05M $K_3Fe(CN)_6$ +0.05M $K_4Fe(CN)_6$ | **44.1** | **25.3** | This work |

**Supporting References**